\documentclass[twocolumn,traditabstract,longauth]{aa}

\usepackage[nonamebreak]{natbib}
\usepackage[stable]{footmisc}
\usepackage{amsmath}
\usepackage{txfonts}
\usepackage{placeins}
\bibpunct{(}{)}{;}{a}{}{,}
\usepackage{graphicx}
\usepackage{epstopdf}
\usepackage{ifthen}
\usepackage{overpic}
\usepackage{rotating}
\usepackage{subfigure}
\usepackage{epsfig}
\usepackage{multirow}
\usepackage{array}
\usepackage{color}
\usepackage[usernames,dvipsnames]{xcolor}
\usepackage{psfrag}
\usepackage[T1]{fontenc}
\usepackage{url}

\newcommand{\hi}{\ion{H}{i}} 
 
\newcommand{\healpix}{{\tt HEALPix}}
\newcommand{\planck}{\textit{Planck}}

\def\clee{$C_\ell^{EE}$}
\def\clbb{$C_\ell^{BB}$}

\def\ben{\begin{enumerate}}
\def\een{\end{enumerate}}
\def\bi{\begin{itemize}}
\def\ei{\end{itemize}}
\def\be{\begin{equation}}
\def\ee{\end{equation}}
\def\bea{\begin{eqnarray}}
\def\eea{\end{eqnarray}}

\newcommand{\StokesI}{I}                    
\newcommand{\StokesQ}{Q}                    
\newcommand{\StokesU}{U}                    

\newcommand{\polangsky}{\gamma}             

\def\setsymbol#1#2{\expandafter\def\csname #1\endcsname{#2}}
\def\getsymbol#1{\csname #1\endcsname}

\def\Planck{\textit{Planck}}

\newbox\tablebox    \newdimen\tablewidth
\def\leaderfil{\leaders\hbox to 5pt{\hss.\hss}\hfil}
\def\tablenote#1 #2\par{\begingroup \parindent=0.8em
    \abovedisplayshortskip=0pt\belowdisplayshortskip=0pt
    \noindent
    $$\hss\vbox{\hsize\tablewidth \hangindent=\parindent \hangafter=1 \noindent
    \hbox to \parindent{$^#1$\hss}\strut#2\strut\par}\hss$$
    \endgroup}

\def\L2{\ifmmode L_2\else $L_2$\fi}

\def\DeltaT{\ifmmode \Delta T\else $\Delta T$\fi}
\def\deltat{\ifmmode \Delta t\else $\Delta t$\fi}
\def\fknee{\ifmmode f_{\rm knee}\else $f_{\rm knee}$\fi}
\def\Fmax{\ifmmode F_{\rm max}\else $F_{\rm max}$\fi}
\def\solar{\ifmmode{\rm M}_{\mathord\odot}\else${\rm M}_{\mathord\odot}$\fi}
\def\Msolar{\ifmmode{\rm M}_{\mathord\odot}\else${\rm M}_{\mathord\odot}$\fi}
\def\Lsolar{\ifmmode{\rm L}_{\mathord\odot}\else${\rm L}_{\mathord\odot}$\fi}
\def\inv{\ifmmode^{-1}\else$^{-1}$\fi}
\def\mo{\ifmmode^{-1}\else$^{-1}$\fi}
\def\sup#1{\ifmmode ^{\rm #1}\else $^{\rm #1}$\fi}
\def\expo#1{\ifmmode \times 10^{#1}\else $\times 10^{#1}$\fi}
\def\,{\thinspace}
\def\lsim{\mathrel{\raise .4ex\hbox{\rlap{$<$}\lower 1.2ex\hbox{$\sim$}}}}
\def\gsim{\mathrel{\raise .4ex\hbox{\rlap{$>$}\lower 1.2ex\hbox{$\sim$}}}}

\def\simprop{\mathrel{\raise .4ex\hbox{\rlap{$\propto$}\lower 1.2ex\hbox{$\sim$}}}}
\def\deg{\ifmmode^\circ\else$^\circ$\fi}
\def\pdeg{\ifmmode $\setbox0=\hbox{$^{\circ}$}\rlap{\hskip.11\wd0 .}$^{\circ}
          \else \setbox0=\hbox{$^{\circ}$}\rlap{\hskip.11\wd0 .}$^{\circ}$\fi}
\def\arcs{\ifmmode {^{\scriptstyle\prime\prime}}
          \else $^{\scriptstyle\prime\prime}$\fi}
\def\arcm{\ifmmode {^{\scriptstyle\prime}}
          \else $^{\scriptstyle\prime}$\fi}
\newdimen\sa  \newdimen\sb
\def\parcs{\sa=.07em \sb=.03em
     \ifmmode \hbox{\rlap{.}}^{\scriptstyle\prime\kern -\sb\prime}\hbox{\kern -\sa}
     \else \rlap{.}$^{\scriptstyle\prime\kern -\sb\prime}$\kern -\sa\fi}
\def\parcm{\sa=.08em \sb=.03em
     \ifmmode \hbox{\rlap{.}\kern\sa}^{\scriptstyle\prime}\hbox{\kern-\sb}
     \else \rlap{.}\kern\sa$^{\scriptstyle\prime}$\kern-\sb\fi}
\def\ra[#1 #2 #3.#4]{#1\sup{h}#2\sup{m}#3\sup{s}\llap.#4}
\def\dec[#1 #2 #3.#4]{#1\deg#2\arcm#3\arcs\llap.#4}
\def\deco[#1 #2 #3]{#1\deg#2\arcm#3\arcs}
\def\rra[#1 #2]{#1\sup{h}#2\sup{m}}

\def\dots{\relax\ifmmode \ldots\else $\ldots$\fi}
\def\WHzsr{\ifmmode $W\,Hz\mo\,sr\mo$\else W\,Hz\mo\,sr\mo\fi}
\def\mHz{\ifmmode $\,mHz$\else \,mHz\fi}
\def\GHz{\ifmmode $\,GHz$\else \,GHz\fi}
\def\mKs{\ifmmode $\,mK\,s$^{1/2}\else \,mK\,s$^{1/2}$\fi}
\def\muKs{\ifmmode \,\mu$K\,s$^{1/2}\else \,$\mu$K\,s$^{1/2}$\fi}
\def\muKRJs{\ifmmode \,\mu$K$_{\rm RJ}$\,s$^{1/2}\else \,$\mu$K$_{\rm RJ}$\,s$^{1/2}$\fi}
\def\muKHz{\ifmmode \,\mu$K\,Hz$^{-1/2}\else \,$\mu$K\,Hz$^{-1/2}$\fi}
\def\MJysr{\ifmmode \,$MJy\,sr\mo$\else \,MJy\,sr\mo\fi}
\def\MJysrmK{\ifmmode \,$MJy\,sr\mo$\,mK$_{\rm CMB}\mo\else \,MJy\,sr\mo\,mK$_{\rm CMB}\mo$\fi}
\def\microns{\ifmmode \,\mu$m$\else \,$\mu$m\fi}

\def\muK{\ifmmode \,\mu$K$\else \,$\mu$\hbox{K}\fi}
\def\microK{\ifmmode \,\mu$K$\else \,$\mu$\hbox{K}\fi}
\def\muW{\ifmmode \,\mu$W$\else \,$\mu$\hbox{W}\fi}
\def\kms{\ifmmode $\,km\,s$^{-1}\else \,km\,s$^{-1}$\fi}
\def\kmsMpc{\ifmmode $\,\kms\,Mpc\mo$\else \,\kms\,Mpc\mo\fi}
\providecommand{\sorthelp}[1]{}

\begin{document}
\author{\small
Planck Collaboration: N.~Aghanim\inst{51}
\and
M.~I.~R.~Alves\inst{82, 9, 51}
\and
D.~Arzoumanian\inst{51, 64}
\and
J.~Aumont\inst{51}
\and
C.~Baccigalupi\inst{74}
\and
M.~Ballardini\inst{26, 42, 45}
\and
A.~J.~Banday\inst{82, 9}
\and
R.~B.~Barreiro\inst{56}
\and
N.~Bartolo\inst{25, 57}
\and
S.~Basak\inst{74}
% ...
}
\institute{\small
APC, AstroParticule et Cosmologie, Universit\'{e} Paris Diderot, CNRS/IN2P3, CEA/lrfu, Observatoire de Paris, Sorbonne Paris Cit\'{e}, 10, rue Alice Domon et L\'{e}onie Duquet, 75205 Paris Cedex 13, France\goodbreak
\and
Aalto University Mets\"{a}hovi Radio Observatory and Dept of Radio Science and Engineering, P.O. Box 13000, FI-00076 AALTO, Finland\goodbreak
\and
African Institute for Mathematical Sciences, 6-8 Melrose Road, Muizenberg, Cape Town, South Africa\goodbreak
\and
Agenzia Spaziale Italiana Science Data Center, Via del Politecnico snc, 00133, Roma, Italy\goodbreak
\and
Aix Marseille Universit\'{e}, CNRS, LAM (Laboratoire d'Astrophysique de Marseille) UMR 7326, 13388, Marseille, France\goodbreak
\and
Astrophysics Group, Cavendish Laboratory, University of Cambridge, J J Thomson Avenue, Cambridge CB3 0HE, U.K.\goodbreak
\and
Astrophysics \& Cosmology Research Unit, School of Mathematics, Statistics \& Computer Science, University of KwaZulu-Natal, Westville Campus, Private Bag X54001, Durban 4000, South Africa\goodbreak
\and
CITA, University of Toronto, 60 St. George St., Toronto, ON M5S 3H8, Canada\goodbreak
\and
CNRS, IRAP, 9 Av. colonel Roche, BP 44346, F-31028 Toulouse cedex 4, France\goodbreak
\and
California Institute of Technology, Pasadena, California, U.S.A.\goodbreak
\and
Computational Cosmology Center, Lawrence Berkeley National Laboratory, Berkeley, California, U.S.A.\goodbreak
\and
Consejo Superior de Investigaciones Cient\'{\i}ficas (CSIC), Madrid, Spain\goodbreak
\and
DTU Space, National Space Institute, Technical University of Denmark, Elektrovej 327, DK-2800 Kgs. Lyngby, Denmark\goodbreak
\and
D\'{e}partement de Physique Th\'{e}orique, Universit\'{e} de Gen\`{e}ve, 24, Quai E. Ansermet,1211 Gen\`{e}ve 4, Switzerland\goodbreak
\and
Departamento de Astrof\'{i}sica, Universidad de La Laguna (ULL), E-38206 La Laguna, Tenerife, Spain\goodbreak
\and
Departamento de F\'{\i}sica, Universidad de Oviedo, Avda. Calvo Sotelo s/n, Oviedo, Spain\goodbreak
\and
Department of Astrophysics/IMAPP, Radboud University Nijmegen, P.O. Box 9010, 6500 GL Nijmegen, The Netherlands\goodbreak
\and
Department of Physics \& Astronomy, University of British Columbia, 6224 Agricultural Road, Vancouver, British Columbia, Canada\goodbreak
\and
Department of Physics and Astronomy, Dana and David Dornsife College of Letter, Arts and Sciences, University of Southern California, Los Angeles, CA 90089, U.S.A.\goodbreak
\and
Department of Physics and Astronomy, University College London, London WC1E 6BT, U.K.\goodbreak
\and
Department of Physics, Gustaf H\"{a}llstr\"{o}min katu 2a, University of Helsinki, Helsinki, Finland\goodbreak
\and
Department of Physics, Princeton University, Princeton, New Jersey, U.S.A.\goodbreak
\and
Department of Physics, University of California, Santa Barbara, California, U.S.A.\goodbreak
\and
Department of Physics, University of Illinois at Urbana-Champaign, 1110 West Green Street, Urbana, Illinois, U.S.A.\goodbreak
\and
Dipartimento di Fisica e Astronomia G. Galilei, Universit\`{a} degli Studi di Padova, via Marzolo 8, 35131 Padova, Italy\goodbreak
\and
Dipartimento di Fisica e Astronomia, Alma Mater Studiorum, Universit\`{a} degli Studi di Bologna, Viale Berti Pichat 6/2, I-40127, Bologna, Italy\goodbreak
\and
Dipartimento di Fisica e Scienze della Terra, Universit\`{a} di Ferrara, Via Saragat 1, 44122 Ferrara, Italy\goodbreak
\and
Dipartimento di Fisica, Universit\`{a} La Sapienza, P. le A. Moro 2, Roma, Italy\goodbreak
\and
Dipartimento di Fisica, Universit\`{a} degli Studi di Milano, Via Celoria, 16, Milano, Italy\goodbreak
\and
Dipartimento di Fisica, Universit\`{a} degli Studi di Trieste, via A. Valerio 2, Trieste, Italy\goodbreak
\and
Dipartimento di Matematica, Universit\`{a} di Roma Tor Vergata, Via della Ricerca Scientifica, 1, Roma, Italy\goodbreak
\and
Discovery Center, Niels Bohr Institute, Copenhagen University, Blegdamsvej 17, Copenhagen, Denmark\goodbreak
\and
European Space Agency, ESAC, Planck Science Office, Camino bajo del Castillo, s/n, Urbanizaci\'{o}n Villafranca del Castillo, Villanueva de la Ca\~{n}ada, Madrid, Spain\goodbreak
\and
European Space Agency, ESTEC, Keplerlaan 1, 2201 AZ Noordwijk, The Netherlands\goodbreak
\and
Gran Sasso Science Institute, INFN, viale F. Crispi 7, 67100 L'Aquila, Italy\goodbreak
\and
HGSFP and University of Heidelberg, Theoretical Physics Department, Philosophenweg 16, 69120, Heidelberg, Germany\goodbreak
\and
Haverford College Astronomy Department, 370 Lancaster Avenue, Haverford, Pennsylvania, U.S.A.\goodbreak
\and
Helsinki Institute of Physics, Gustaf H\"{a}llstr\"{o}min katu 2, University of Helsinki, Helsinki, Finland\goodbreak
\and
INAF - Osservatorio Astronomico di Padova, Vicolo dell'Osservatorio 5, Padova, Italy\goodbreak
\and
INAF - Osservatorio Astronomico di Roma, via di Frascati 33, Monte Porzio Catone, Italy\goodbreak
\and
INAF - Osservatorio Astronomico di Trieste, Via G.B. Tiepolo 11, Trieste, Italy\goodbreak
\and
INAF/IASF Bologna, Via Gobetti 101, Bologna, Italy\goodbreak
\and
INAF/IASF Milano, Via E. Bassini 15, Milano, Italy\goodbreak
\and
INFN - CNAF, viale Berti Pichat 6/2, 40127 Bologna, Italy\goodbreak
\and
INFN, Sezione di Bologna, viale Berti Pichat 6/2, 40127 Bologna, Italy\goodbreak
\and
INFN, Sezione di Ferrara, Via Saragat 1, 44122 Ferrara, Italy\goodbreak
\and
INFN, Sezione di Roma 1, Universit\`{a} di Roma Sapienza, Piazzale Aldo Moro 2, 00185, Roma, Italy\goodbreak
\and
INFN, Sezione di Roma 2, Universit\`{a} di Roma Tor Vergata, Via della Ricerca Scientifica, 1, Roma, Italy\goodbreak
\and
INFN/National Institute for Nuclear Physics, Via Valerio 2, I-34127 Trieste, Italy\goodbreak
\and
Imperial College London, Astrophysics group, Blackett Laboratory, Prince Consort Road, London, SW7 2AZ, U.K.\goodbreak
\and
Institut d'Astrophysique Spatiale, CNRS, Univ. Paris-Sud, Universit\'{e} Paris-Saclay, B\^{a}t. 121, 91405 Orsay cedex, France\goodbreak
\and
Institut d'Astrophysique de Paris, CNRS (UMR7095), 98 bis Boulevard Arago, F-75014, Paris, France\goodbreak
\and
Institute of Astronomy, University of Cambridge, Madingley Road, Cambridge CB3 0HA, U.K.\goodbreak
\and
Institute of Theoretical Astrophysics, University of Oslo, Blindern, Oslo, Norway\goodbreak
\and
Instituto de Astrof\'{\i}sica de Canarias, C/V\'{\i}a L\'{a}ctea s/n, La Laguna, Tenerife, Spain\goodbreak
\and
Instituto de F\'{\i}sica de Cantabria (CSIC-Universidad de Cantabria), Avda. de los Castros s/n, Santander, Spain\goodbreak
\and
Istituto Nazionale di Fisica Nucleare, Sezione di Padova, via Marzolo 8, I-35131 Padova, Italy\goodbreak
\and
Jet Propulsion Laboratory, California Institute of Technology, 4800 Oak Grove Drive, Pasadena, California, U.S.A.\goodbreak
\and
Jodrell Bank Centre for Astrophysics, Alan Turing Building, School of Physics and Astronomy, The University of Manchester, Oxford Road, Manchester, M13 9PL, U.K.\goodbreak
\and
Kavli Institute for Cosmological Physics, University of Chicago, Chicago, IL 60637, USA\goodbreak
\and
Kavli Institute for Cosmology Cambridge, Madingley Road, Cambridge, CB3 0HA, U.K.\goodbreak
\and
LAL, Universit\'{e} Paris-Sud, CNRS/IN2P3, Orsay, France\goodbreak
\and
LERMA, CNRS, Observatoire de Paris, 61 Avenue de l'Observatoire, Paris, France\goodbreak
\and
Laboratoire AIM, IRFU/Service d'Astrophysique - CEA/DSM - CNRS - Universit\'{e} Paris Diderot, B\^{a}t. 709, CEA-Saclay, F-91191 Gif-sur-Yvette Cedex, France\goodbreak
\and
Laboratoire Traitement et Communication de l'Information, CNRS (UMR 5141) and T\'{e}l\'{e}com ParisTech, 46 rue Barrault F-75634 Paris Cedex 13, France\goodbreak
\and
Laboratoire de Physique Subatomique et Cosmologie, Universit\'{e} Grenoble-Alpes, CNRS/IN2P3, 53, rue des Martyrs, 38026 Grenoble Cedex, France\goodbreak
\and
Laboratoire de Physique Th\'{e}orique, Universit\'{e} Paris-Sud 11 \& CNRS, B\^{a}timent 210, 91405 Orsay, France\goodbreak
\and
Lawrence Berkeley National Laboratory, Berkeley, California, U.S.A.\goodbreak
\and
Max-Planck-Institut f\"{u}r Astrophysik, Karl-Schwarzschild-Str. 1, 85741 Garching, Germany\goodbreak
\and
Mullard Space Science Laboratory, University College London, Surrey RH5 6NT, U.K.\goodbreak
\and
Nicolaus Copernicus Astronomical Center, Bartycka 18, 00-716 Warsaw, Poland\goodbreak
\and
Niels Bohr Institute, Copenhagen University, Blegdamsvej 17, Copenhagen, Denmark\goodbreak
\and
Nordita (Nordic Institute for Theoretical Physics), Roslagstullsbacken 23, SE-106 91 Stockholm, Sweden\goodbreak
\and
SISSA, Astrophysics Sector, via Bonomea 265, 34136, Trieste, Italy\goodbreak
\and
School of Physics and Astronomy, University of Nottingham, Nottingham NG7 2RD, U.K.\goodbreak
\and
Simon Fraser University, Department of Physics, 8888 University Drive, Burnaby BC, Canada\goodbreak
\and
Sorbonne Universit\'{e}-UPMC, UMR7095, Institut d'Astrophysique de Paris, 98 bis Boulevard Arago, F-75014, Paris, France\goodbreak
\and
Space Sciences Laboratory, University of California, Berkeley, California, U.S.A.\goodbreak
\and
Sub-Department of Astrophysics, University of Oxford, Keble Road, Oxford OX1 3RH, U.K.\goodbreak
\and
The Oskar Klein Centre for Cosmoparticle Physics, Department of Physics,Stockholm University, AlbaNova, SE-106 91 Stockholm, Sweden\goodbreak
\and
UPMC Univ Paris 06, UMR7095, 98 bis Boulevard Arago, F-75014, Paris, France\goodbreak
\and
Universit\'{e} de Toulouse, UPS-OMP, IRAP, F-31028 Toulouse cedex 4, France\goodbreak
% ...
}

\title{\textit{Planck} intermediate results. XLIV.\\
The structure of the Galactic magnetic field\\
from dust polarization maps of the southern Galactic cap}

\abstract{Using data from the \Planck\ satellite, we study the statistical
properties of interstellar dust polarization at high Galactic latitudes.
Our aim is to advance the understanding of the magnetized interstellar medium
(ISM), and to provide a modelling framework of the polarized dust foreground
for use in cosmic microwave background (CMB) component-separation procedures.
Focusing on the southern Galactic cap ($b<-60\deg$), we examine the
Stokes $I$, $Q$, and $U$ maps at 353\,GHz, and particularly the statistical
distribution of the polarization fraction ($p$) and angle ($\psi$), in order
to characterize the ordered and turbulent components of the Galactic magnetic
field (GMF) in the solar neighbourhood.  The $Q$ and $U$ maps show patterns at
large angular scales, which we relate to the mean orientation of the GMF
towards Galactic coordinates $(l_0,b_0)=(70\deg\pm5\deg,24\deg\pm5\deg)$.
The histogram of the observed $p$ values shows a wide dispersion up to
25\,\%.  The histogram of $\psi$ has a standard deviation of
$12\deg$ about the regular pattern expected from the ordered GMF.
We build a phenomenological model that connects the distributions of $p$ and $\psi$ to a statistical description of the 
turbulent component of the GMF, assuming
a uniform effective polarization fraction ($p_0$) of dust emission.
To compute the Stokes parameters, we approximate the integration
along the line of sight (LOS) as a sum over a set of $N$ independent
polarization layers, in each of which the turbulent component of the
GMF is obtained from Gaussian realizations of a power-law power spectrum. 
We are able to reproduce the observed $p$ and $\psi$ distributions using:
a $p_0$ value of $26\,\%$; a ratio of 0.9 between the strengths of the
turbulent and mean components of the GMF; and a small value of $N$. The mean
value of $p$ (inferred from the fit of the large-scale patterns in the Stokes
maps) is $12\pm1\,$\%. We relate the polarization layers to the density
structure and to the correlation length of the GMF along the LOS.
We stress the simplicity of our model (involving only a few parameters),
which can be easily computed on the celestial sphere to produce
simulated maps of dust polarization, and thereby to assess component-separation
approaches in CMB experiments.}


\titlerunning{The local structure of the Galactic magnetic field}
\authorrunning{Planck Collaboration}
\maketitle

\section{Introduction}\label{sec:intro}
Interstellar magnetic fields are tied to the interstellar gas. 
Together with cosmic rays they form a dynamical system that is an important
(but debated) facet of the physics of galaxies.  Magnetic fields play a pivotal
role, because they control the density and distribution of cosmic rays, and they
act on the dynamics if the gas.  Much of the physics involved in this interplay
is encoded in the structure of interstellar magnetic fields. Observations of
synchrotron emission and its polarization, as well as Faraday rotation and dust
polarization, provide the means to characterize the structure of magnetic
fields within galaxies \citep{Haverkorn15,Lazarian16,Beck16}. 

Since dust grains are mixed with interstellar gas, dust polarization data are
well suited to investigate the physical coupling between the gas dynamics and
the magnetic field structure, in other words to characterize 
magnetohydrodynamical (MHD) turbulence in the interstellar medium
\citep[ISM;][]{Brandenburg13,Falceta14}.
Anisotropic dust grains tend to align with their longer axes perpendicular to
the local magnetic field, and thus their emission is polarized 
perpendicular to the magnetic field projection on the plane of the sky (POS). 
The polarization fraction, $p$, the ratio between the polarized and
total intensities of dust thermal emission, depends on the dust polarization
properties and the grain alignment efficiency, but also on
the structure of the magnetic field  \citep{Lazarian07}.
Thus, information on the magnetic field structure is encoded in the Stokes $Q$
and $U$ maps, as well as in the polarization angle $\psi$ and fraction $p$.

For a long time, observations of dust polarization from the diffuse ISM were
limited to stellar polarization data available for a
discrete set of lines of sight \citep[LOS;][]{Heiles00}.
The \Planck\footnote{\Planck\ (\url{http://www.esa.int/Planck}) is a project of
the European Space Agency (ESA) with instruments provided by two scientific
consortia funded by ESA member states and led by Principal Investigators from
France and Italy, telescope reflectors provided through a collaboration between
ESA and a scientific consortium led and funded by Denmark, and additional
contributions from NASA (USA).} data opened a new perspective on this topic.
For the first time, we have maps of the dust polarization in emission over
the full sky \citep{planck2014-a01}. 
The \Planck\ maps greatly supersede, in sensitivity and statistical power,
the data available from earlier ground-based and balloon-borne observations
\citep[e.g.,][]{Benoit04,Ponthieu05,Ward09,Koch10,Poidevin14,Matthews14}.

Several studies have already used the \Planck\ data to investigate the link 
between the dust polarization maps and the structure of the Galactic
magnetic field (GMF).  \citet{planck2014-XIX} presented the first analysis of
the polarized sky as seen at 353\,GHz (the most sensitive \Planck\
channel for polarized thermal dust emission), focusing on the statistics of 
$p$ and $\psi$.  The comparison with synthetic polarized emission maps,
computed from simulations of anisotropic MHD turbulence, shows that the
turbulent structure of the GMF is able to reproduce the main statistical 
properties of $p$ and $\psi$ in nearby molecular clouds \citep{planck2014-XX}.
This comparison shows that the mean orientation of the GMF with respect 
to the LOS plays a major role in the quantitative analysis of these statistical
properties.  An important result is that in the diffuse ISM, the filamentary
structure of matter is observed to be statistically aligned with 
the GMF \citep{McClure06,Clark13,planck2014-XXXII,Kalberla16}.

The spatial structure of the polarization angle has been characterized in \citet{planck2014-XIX} using the angle
dispersion function $\mathcal{S}$. The map of $\mathcal{S}$ highlights long,
narrow structures of high $\mathcal{S}$ that trace abrupt changes of $\psi$
at the interfaces between extended areas within which the polarization angle
is ordered.  \citet{Falgarone15} found a correlation between the structures in
$\mathcal{S}$ and large velocity shears in incompressible magnetized
turbulence.  The structures seen in the \Planck\ data bear a morphological
resemblance to features associated with Faraday rotation in gradient maps
of polarized synchrotron emission \citep{Gaensler11,Iacobelli14}, which have
been related to fluctuations in the GMF and in the ionized gas density in MHD
turbulence \citep{Burkhart12}.
Filamentary structures in rotation measure synthesis maps from LOFAR
(the Low-Frequency Array) data \citep{Jelic15} have been shown to be
correlated with the GMF orientation inferred from the \Planck\ dust
polarization \citep{Zaroubi15}.  At microwave frequencies, the dust
polarization has been demonstrated to be correlated with synchrotron
polarization, free from Faraday rotation \citep{planck2014-XXII,Choi15}.
Both emission processes trace the same GMF, but the correlation is not
one-to-one due to the difference in the distribution of dust and relativistic
electrons in the Galaxy.  \citet{Jaffe13} and \citet{Planck2016-XLII} described
the difficulties faced when trying to reproduce the \Planck\ dust polarization
data with existing models of the large-scale GMF
\citep{Jaffe10,Sun10,Jansson12}, which are mainly constrained by
synchrotron emission and Faraday rotation measures.

\begin{figure*}
\hspace{-0.8cm}
\begin{tabular}{r c l}
\includegraphics[width=6.cm]{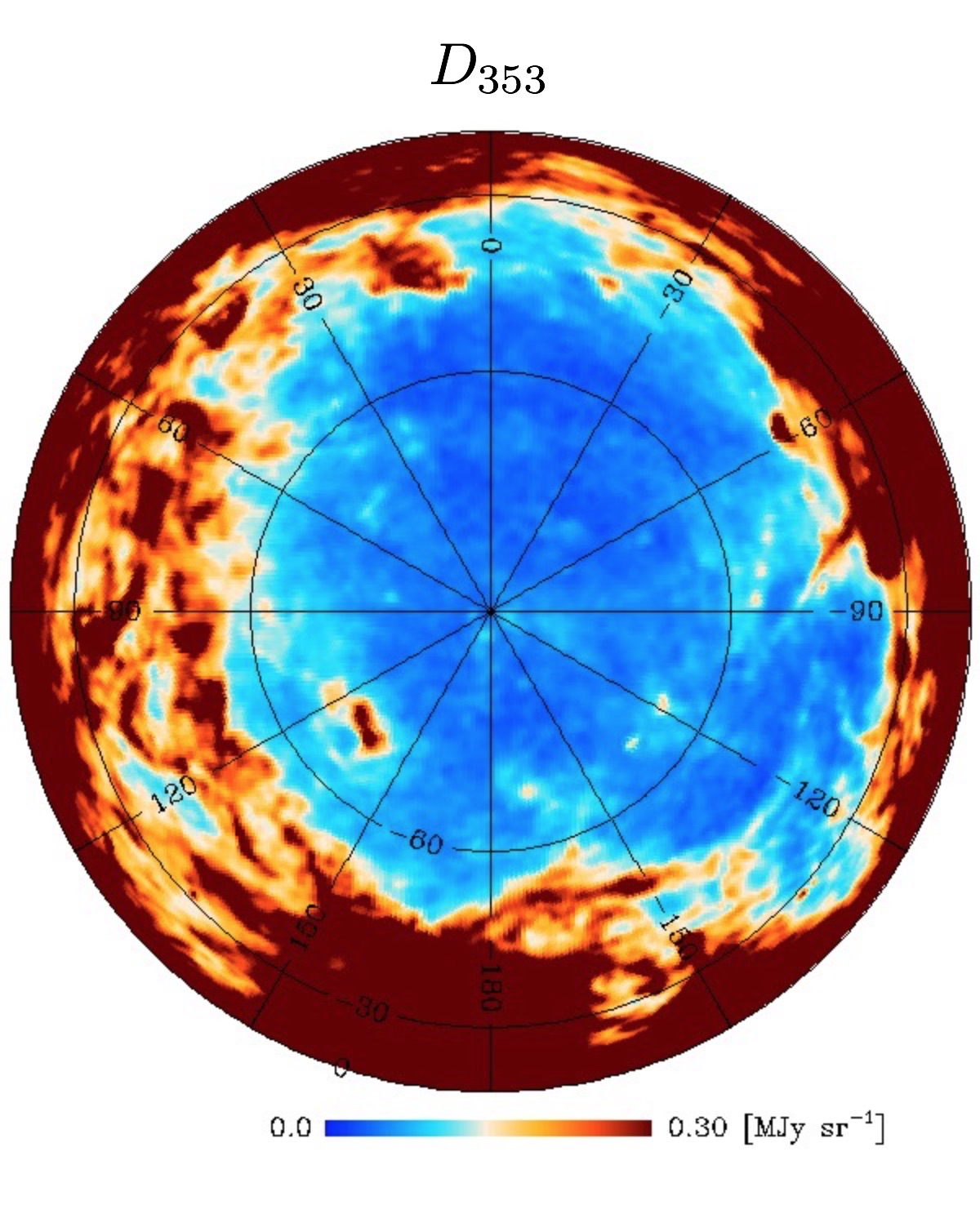}
 & \includegraphics[width=6.cm]{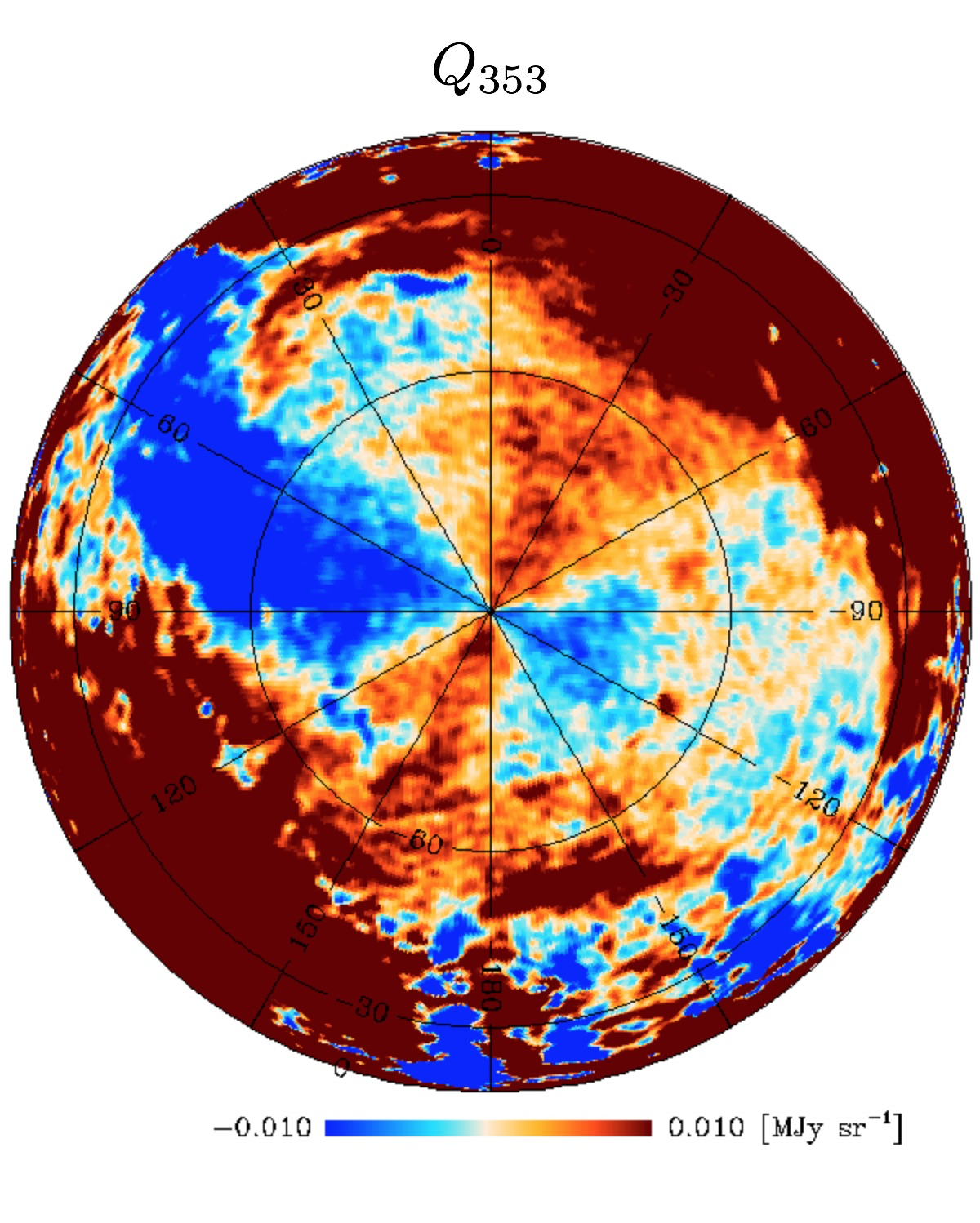}
 & \includegraphics[width=6.cm]{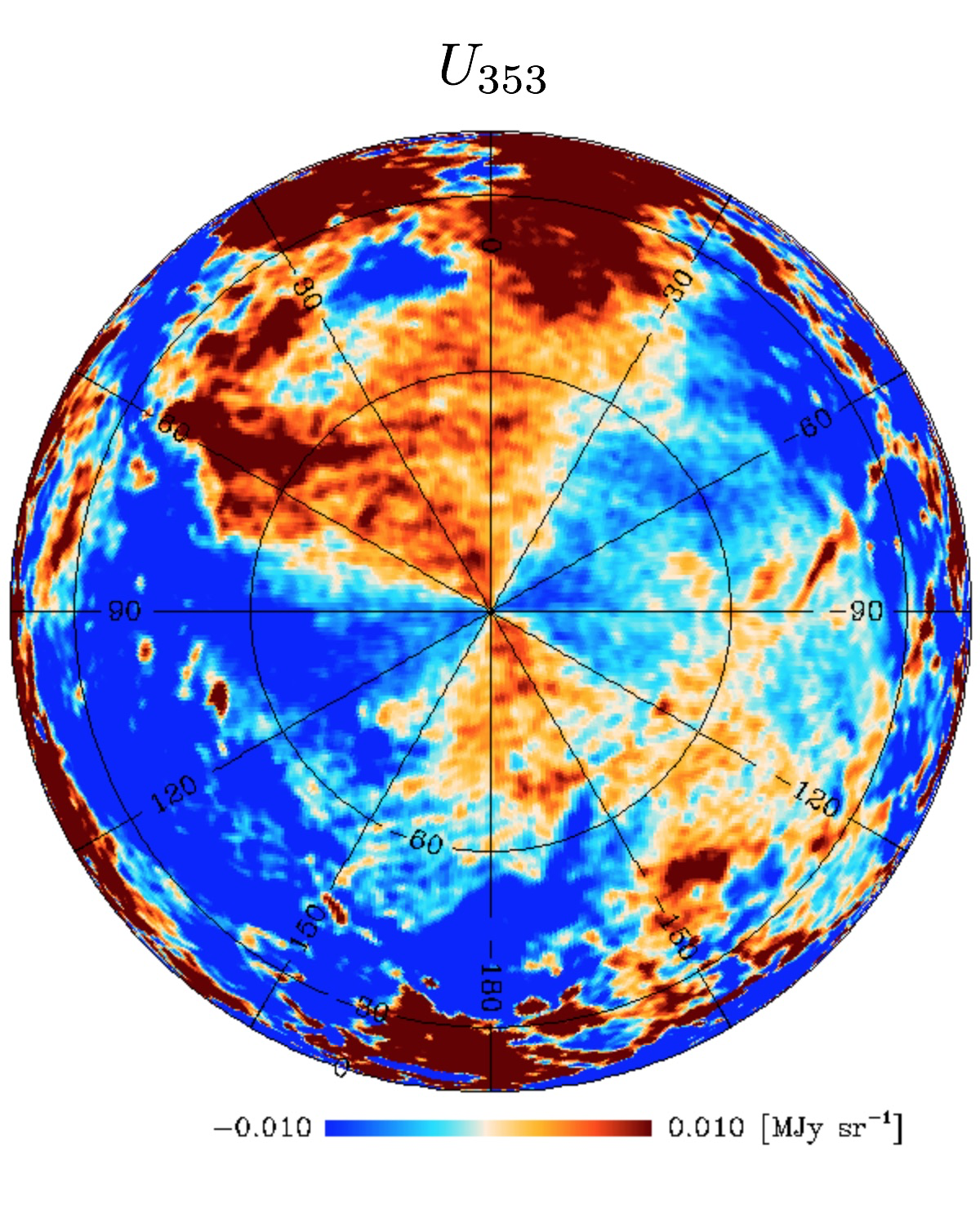}\\
\end{tabular}
\caption[]{Orthographic projections centred on the south Galactic
  pole of the \planck\ dust emission
  intensity, $D_{353}$ ({\it left}), and the Stokes
  $Q_{353}$ ({\it centre}) and $U_{353}$ ({\it right}) maps, at
  353\,GHz. A grid of Galactic coordinates is included, labelled in degrees.
  East is on the left of the maps and the west on the right. Note that
  the $U_{353}$ map is in the \healpix\ (or CMB) polarization convention,
  which corresponds to $-U_{353}$ in the IAU convention.}
\label{fig:data353}
\end{figure*}


The GMF structure is also relevant for the modelling of polarized Galactic
foregrounds in analyses of the CMB.  Thermal emission from Galactic dust
is the main polarized foreground at frequencies above 100\,GHz
\citep{planck2014-a12}.  \citet{planck2014-XXX} presented the polarized dust
angular power spectra \clee\ and \clbb, 
providing cosmologists with a characterization of the 
dust foreground to CMB polarization.  \citet{planck2015-XXXVIII} showed that 
the correlation between the filamentary structure of matter and the
GMF orientation may account for the $E$ and $B$ asymmetry, 
as well as the $TE$  correlation, 
reported in the analysis of the power spectra of the \planck\ 353\,GHz polarization maps. 

Within this broad context, the motivations and objectives of this paper are
twofold. First, we extend the analysis of the \Planck\ dust polarization maps
to the high Galactic latitude sky that was masked in the \citet{planck2014-XIX}
analysis, because of residual systematic errors in the data.  The polarization
maps at 353\,GHz \citep{planck2014-a01,planck2014-a09} that have been made
publicly available by the \Planck\
consortium\footnote{\url{http://pla.esac.esa.int}}
are now suitable for such an analysis. 
Second, we introduce a modelling framework that relates the dust
polarization to the GMF structure, its mean orientation, and a statistical
description of its random (turbulent) component. This framework is also a step
towards a modelling tool for the dust polarization, which may be used to assess
component-separation methods in the analysis of CMB polarization
\citep[e.g.,][]{planck2014-a11,planck2014-a12}.

Our data analysis procedure focuses on the southern Galactic cap, the cleanest
part of the sky that is directly relevant to CMB observations, in particular
those carried out with ground-based telescopes from Antarctica and
Chile.\footnote{See \url{http://lambda.gsfc.nasa.gov/product/expt/}}
This is also the part of the sky where the LOS through the Galaxy is the
shortest, and hence is the region best suited to characterize the turbulent
component of the GMF.

The paper is organized as follows. We present the \planck\ data in
Sect.~\ref{sec:data}.  Section~\ref{sec:framework} introduces our model of the
GMF structure in the solar neighbourhood and in Sect.~\ref{sec:ordfield} we
estimate the mean orientation of the GMF in the solar neighbourhood.
In Sect.~\ref{sec:turbfield}, we characterize the turbulent component of the
GMF.  The data analysis is based on a phenomenological model that we discuss
in Sect.~\ref{sec:discussion}, which also contains our future perspectives.
The paper's results are summarized in Sect.~\ref{sec:summary}.
The approximations made to compute the Stokes parameters are presented in
Appendix~\ref{app:approx}.


\section{Data and conventions}\label{sec:data}

We first introduce the data that we will use, discussing the conventions
assumed in the analysis of polarization, and presenting the
polarization parameters determined around the south Galactic pole.

\subsection{Description of the data}\label{ssec:datachar}

The \Planck\ satellite observed the polarized sky in seven frequency bands
from 30 to 353\,GHz \citep{planck2013-p01}.  In this paper, we only use the
data from the High Frequency Instrument \citep[HFI,][]{Lamarre:2010} at the
highest frequency, 353\,GHz, where the dust emission is the brightest. 

We use the publicly available 353\,GHz Stokes $Q$ and $U$ (hereafter,
$\StokesQ_{353}$ and $\StokesU_{353}$) maps (central and right panels in
Fig.~\ref{fig:data353}) and the associated noise maps made with the five
independent consecutive sky surveys of the \Planck\ cryogenic mission. 
We refer to publications by the Planck Collaboration for details of
the processing of HFI data, including mapmaking, photometric calibration, and
photometric uncertainties \citep{planck2014-a01,planck2014-a08,planck2014-a09}.
The $\StokesQ_{353}$ and $\StokesU_{353}$ maps are corrected for spectral
leakage as described in \citet{planck2014-a09}. 
For the dust total intensity at 353\,GHz we use the model map, $D_{353}$,
derived from a modified blackbody fit to the \planck\ data at
$\nu \ge 353\,$GHz, and IRAS at $\lambda = 100\,\mu$m
\citep[][left panel in Fig.~\ref{fig:data353}]{planck2013-p06b}.
The data used in this fit are corrected for zodiacal emission and CMB anisotropies. 
$D_{353}$ has a also a lower noise than the corresponding  353\,GHz Stokes $\StokesI$ \Planck\ map.  
The $\StokesQ_{353}$ and $\StokesU_{353}$ maps are initially constructed with
an effective beamsize of 4\parcm8, and $D_{353}$ at $5\arcmin$. The three maps 
are in \healpix\ format\footnote{\citet{Gorski05}, \tt{http://healpix.sf.net}}
with a pixelization $N_{\rm side} = 2048$. To increase the
signal-to-noise ratio at high Galactic latitudes, we smooth the three maps to
$1^{\circ}$ resolution using a Gaussian approximation to the \Planck\ beam.
We reduce the \healpix\ resolution to $N_{\rm side} = 128$ (30\parcm1 pixels)
after smoothing.  For the polarization maps, we apply the
``\ensuremath{\tt ismoothing}''
routine of \healpix, which decomposes the $\StokesQ$ and $\StokesU$ maps
into $E$ and $B$ maps, applies Gaussian smoothing in harmonic space, 
and transforms the smoothed $E$ and $B$ back into $\StokesQ$ and $\StokesU$
maps at $N_{\rm side}=128$ resolution. 

\subsection{Applied conventions in polarization}\label{ssec:Stokespara}
\begin{figure}
\includegraphics[width=8.5cm]{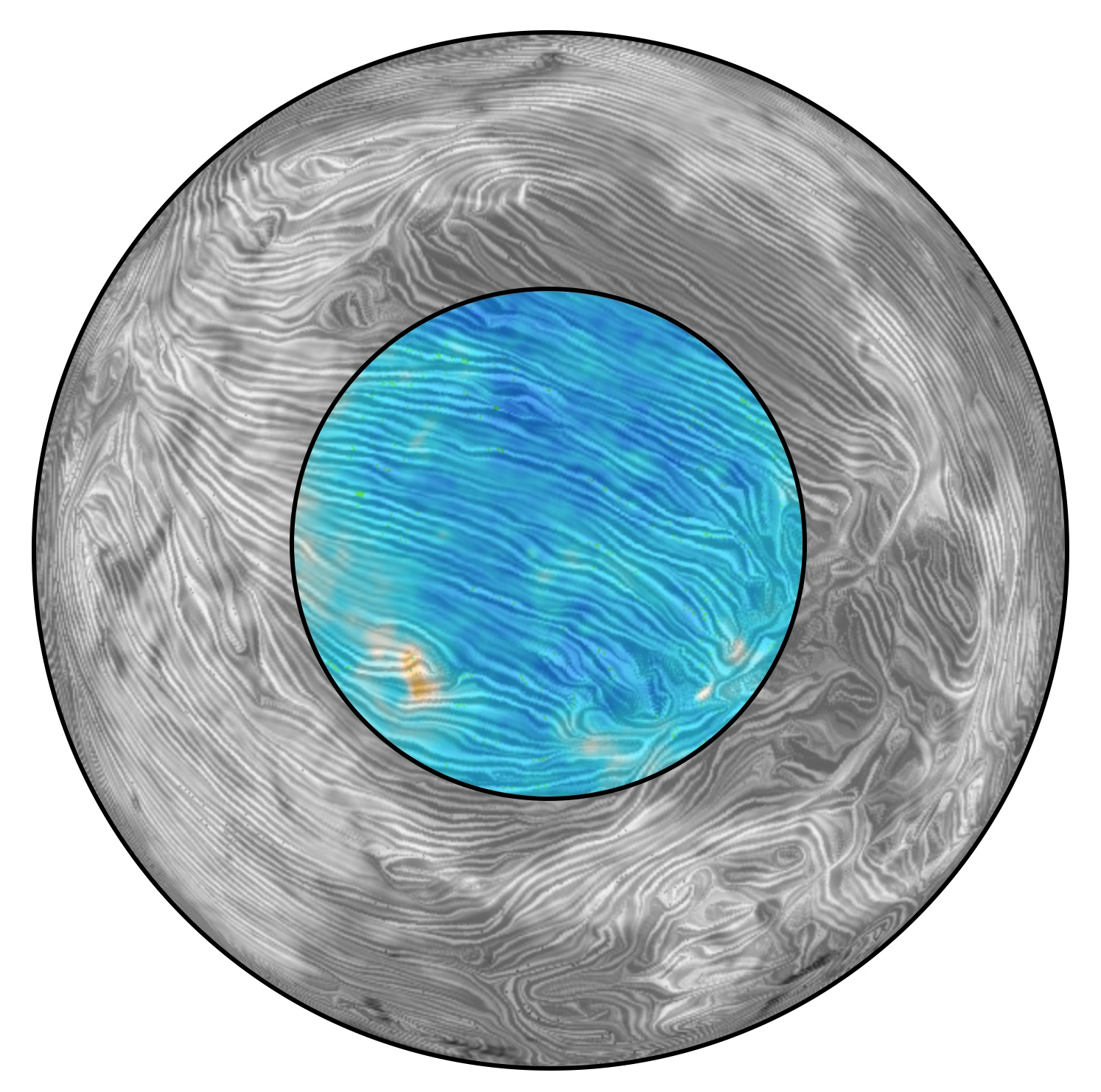} 
\caption[]{\planck\ $D_{353}$ (same as in the left panel of
  Fig.~\ref{fig:data353}) with the ``drapery'' pattern, orthogonal to the
  polarization orientation, produced with the LIC algorithm. The part of the
  sky at $b < -60^\circ$ has been highlighted in colour in this figure. }
\label{fig:lic_image}
\end{figure}

\begin{figure}
\hspace{-0.8cm}
\includegraphics[width=9.5cm]{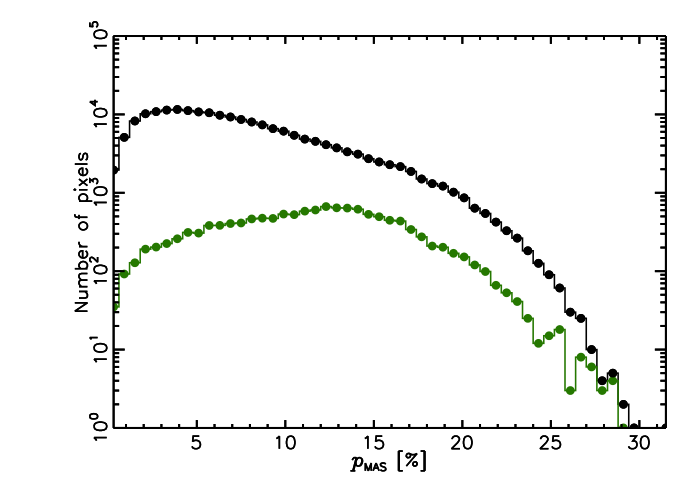}
\caption[]{Histograms of the polarization fraction from the $p_{\rm MAS}$
  debiased estimator (see text). The black histogram shows $p_{\rm MAS}$
  over the whole sky. The green histogram shows $p_{\rm MAS}$ at
  $b < -60^{\circ}$.}
\label{fig:pmas_histo}
\end{figure}


In terms of $Q_{353}$, $U_{353}$, and $D_{353}$, the quantities $p$ and $\psi$,
are defined as
\begin{align}
\label{eq:observables}
 &p = \frac{\sqrt{Q_{353}^2+U_{353}^2}}{D_{353}},\nonumber \\
& \psi= \frac{1}{2}\,{\rm tan}^{-1}\left(-U_{353},Q_{353}\right),
\end{align}
where the minus sign in $\psi$ is needed to change the {\healpix}-format maps
(or ``COSMO convention'' for the FITS keyword {\tt POLCONV})
into the International Astronomical Union (IAU) convention for $\psi$,
measured from the local direction to the north Galactic pole with increasing
positive values towards the east. Moreover, in this paper we use the version
of the inverse tangent function with two signed arguments
to resolve the $\pi$ ambiguity ($\psi $ corresponds to orientations not to directions).

When considering dust polarization, the Stokes parameters for linear
polarization are integral quantities of the optical depth
\citep[see Appendix~\ref{app:approx} and][]{planck2014-XX}.
An empirical expression for $p$ is 
\begin{equation}\label{eq:p_obs}
p = p_{0}\,F \, \cos^2{\gamma},
\end{equation}
where $\gamma$ is the angle between the mean orientation of the GMF and the
POS.  Therefore, the projection factor, $\cos^2 \gamma$, carries information
on the orientation of the GMF with respect to the POS. In particular, dust
polarization vanishes where the GMF points directly towards or away from the
observer.  Hereafter, $p_{0}=p_{\rm dust}\, R$ is the effective dust
polarization fraction, which combines the intrinsic polarization fraction of
dust grains $p_{\rm dust}$ \citep[the ratio between the
polarization and average cross-sections of dust, as defined in][]
{planck2014-XX} and $R$, the Rayleigh reduction factor \citep[related
to the degree of dust grain alignment with the GMF;][]{Greenberg68,Lee85}.
The factor $R$ is equal to 1 for perfect grain alignment.
The factor $F$ accounts for the depolarization due to variations of
the GMF orientation along the LOS and within the beam. 

\begin{figure*}
\begin{tabular}{r l}
\includegraphics[width=8.5cm]{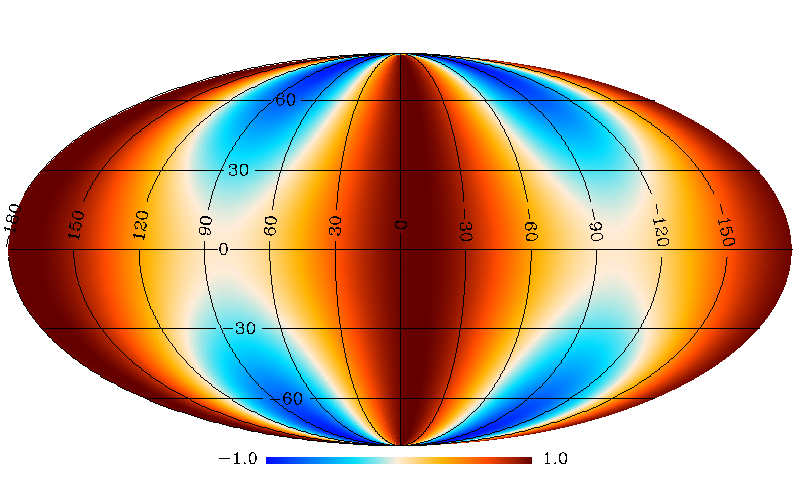}
 & \includegraphics[width=8.5cm]{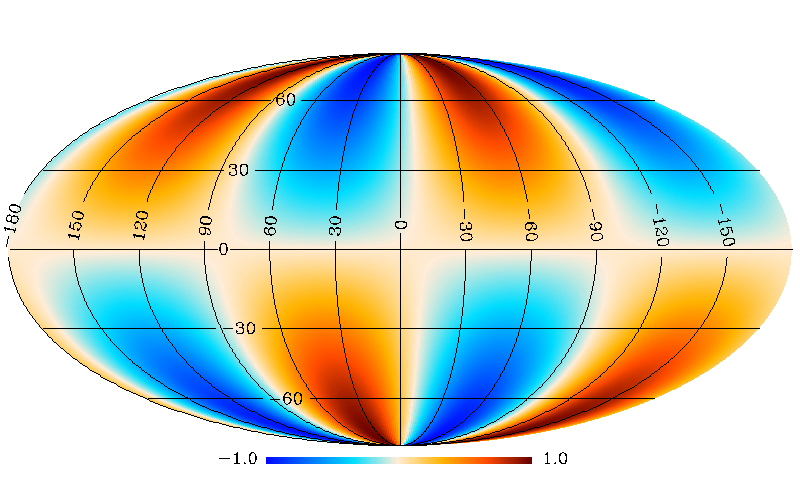}\\
\includegraphics[width=8.5cm]{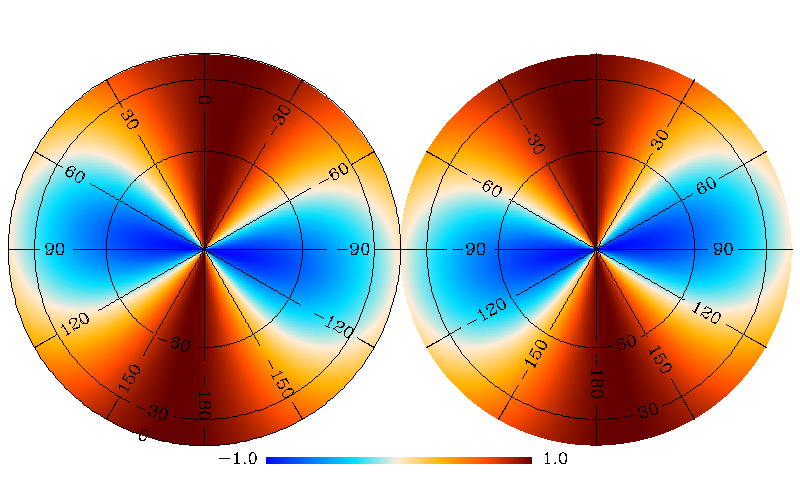}
 & \includegraphics[width=8.5cm]{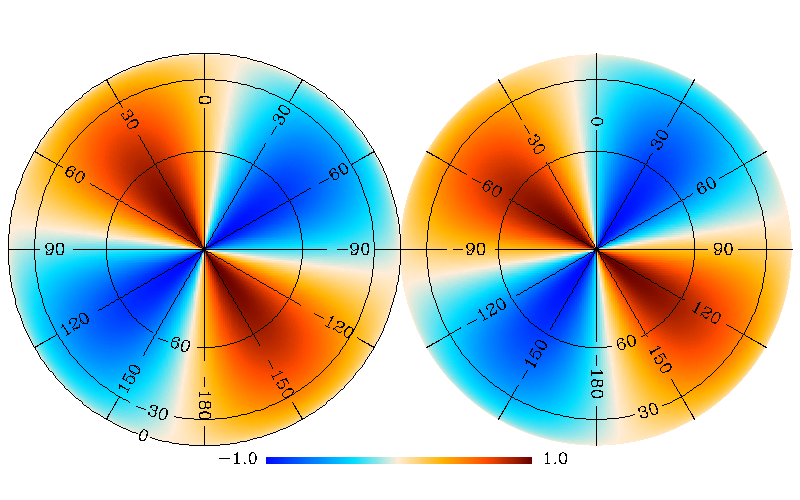}\\
\end{tabular}
\caption[]{Mollweide ({\it top}) and orthographic ({\it bottom})
  projections of the model Stokes parameters, $q_{\rm A}$ ({\it left}) and
  $u_{\rm A}$ ({\it right}), for a uniform direction of the GMF
  towards $(l_0,b_0)=(80^\circ,0^\circ)$; these are roughly the values inferred
  from starlight polarization \citep{Heiles96}. The orthographic projections
  are centred on the Galactic poles. Galactic coordinates in degrees are shown
  on all plots.}
\label{fig:Unimodel}
\end{figure*}

\subsection{Polarization parameters at high Galactic latitudes}
\label{ssec:poldata}

\citet{planck2014-XIX} characterized the polarized
sky at 353\,GHz at low and intermediate Galactic latitudes. Now, with the
maps released in early 2015 \citep{planck2014-a01},
we can extend this analysis to high Galactic latitudes. 
In this work, we focus on the region around the south Galactic pole
(Galactic latitude $b<-60^{\circ}$), which is well suited
to study emission from dust in the diffuse ISM, and directly relevant to
study the dust foreground for CMB polarization. 

We compute $p$ and $\psi$ from the Stokes parameters in
Fig.~\ref{fig:data353} at a resolution of $1^\circ$. Because of the square
of $Q$ and $U$, and the contribution from noise, $p$ cannot be computed
directly from Eq.~(\ref{eq:observables}) at high Galactic latitudes where the
\planck\ signal-to-noise is low. A number of algorithms have been proposed
\citep[e.g., ][]{Montier15} to derive unbiased estimates of $p$; here, we
use the $p_{\rm MAS}$ estimator presented in \citet{Montier14}.

Figure~\ref{fig:lic_image} shows a map of the Planck dust emission
intensity, $D_{353}$, with the ``drapery'' pattern of 
$\psi$, rotated by $\pi/2$, produced with the linear integral 
convolution (LIC) algorithm \citep{Cabral1993} as in \citet{planck2015-XXXV}
and \citet{planck2014-a01}.  This map reveals a high degree of order in $\psi$
at $b < -60^\circ$ (blue region).  Figure~\ref{fig:pmas_histo} shows
histograms of the polarization fraction from the $p_{\rm MAS}$ unbiased
estimator, over the whole sky (black line) and at $b < -60^\circ$ (green line).
Both histograms indicate a wide distribution of $p_{\rm MAS}$, with values up
to 25\,\%; they have comparable dispersions, but they differ for very low
values of $p_{\rm MAS}$.  This difference is due to depolarization from LOS
variations of the GMF orientation on and near the Galactic plane
\citep{planck2014-XIX}.

How do we explain the high $p_{\rm MAS}$ values at high Galactic latitudes
and the observed dispersion in the distribution?  As we will show,
the GMF structure in the solar neighbourhood is essential to consider when
answering this question.


\section{Model framework}
\label{sec:framework}

The polarization of thermal dust emission results from the alignment
of elongated grains with respect to the GMF \citep{Stein66,Hildebrand88}.
Within the hypothesis that grain polarization properties, including alignment,
are homogeneous, the structure of the 
dust polarization sky reflects the structure of the GMF combined with that of
matter. Throughout the paper, we assume that this 
hypothesis applies to the diffuse ISM, where radiative torques provide
a mechanism to efficiently align grains \citep{Dolginov76,Hoang14,Andersson15}.
Our data modelling focusses on the structure of the GMF. This section describes
the model framework (Sect.~\ref{subsec:magnetic_field}) and how we proceed to
fit it to the data (Sect.~\ref{subsec:data_fit}).
 
\subsection{Magnetic field modelling}
\label{subsec:magnetic_field}

We now introduce the framework we use to model the GMF
structure within the solar neighbourhood. The integral equations of the
Stokes $I$, $Q$ and $U$ parameters 
are recalled in Appendix~\ref{app:approx}.

We follow earlier works \citep[e.g.,][]{Chandra1953,Hildebrand2009},
expressing the GMF ($\vec{B}$) as the sum of its mean ($\vec{B}_0$)
and turbulent ($\vec{B}_{\rm t}$) components:
\begin{equation}\label{eq:Bfield} 
\vec{B}=\vec{B}_0+\vec{B}_{\rm t}.
\end{equation}
We introduce and discuss the assumptions we make about each of these two
components. 

Our model aims at describing dust polarization towards the southern Galactic
cap at Galactic latitudes $b \le -60^\circ$. 
We focus on the solar neighbourhood and thereby ignore 
the structure of the GMF on Galaxy-wide scales.  We also ignore the change of
its orientation from the disk to the halo \citep{Haverkorn15}, because dust
emission arises mainly from a thin disk. 
The dust scale height is not measured in the solar neighbourhood,
but modelling of the dust emission from the Milky Way
indicates that the dust scale height at the solar distance to the Galactic
centre is approximately 200\,pc \citep{Drimmel01}.  Observations of the
edge-on spiral galaxy NGC~891, a galaxy analogous to the Milky Way, give
a comparable scale height of around 150\,pc \citep{Bocchio16}.  These estimates
are in agreement with the scale height of the neutral atomic gas in the
Milky Way, inferred from \hi\ observations \citep{Dickey90,Kalberla07}. Hence,
we assume that the vector $\vec{B}_0$ has a fixed orientation, which
represents the mean orientation of the GMF in the solar neighbourhood. 

Radio observations of synchrotron emission and polarization
reveal a wealth of structures down to pc and sub-pc scales
\citep[e.g.,][]{Reich2004,Gaensler11,Iacobelli13,Iacobelli14}, such as
filaments, canals, lenses, and rings, which carry valuable information
about $\vec{B}_{\rm t}$ \citep{Fletcher2006}. \citet{Heiles95} and
\citet{Haverkorn15} reviewed observations that characterize this random
component, concluding that it has a strength 
of about $5\,\mu$G, comparable to that of $\vec{B}_0$.
\citet{Jones92} reached a similar conclusion from stellar polarization data. 

The turbulent component of the GMF is significant. To take it into account, we
follow earlier works \citep[e.g.,][]{Waelkens09,Fauvet11}, modelling each
component of the $\vec{B}_{\rm t}$ vector with Gaussian realizations.
To model dust polarization over the celestial sphere, earlier studies
\citep[e.g.,][]{MAMD2008,Fauvet11,ODea12} computed independent realizations 
of the components of $\vec{B}_{\rm t}$ for each LOS. This approach ignores the
angular coherence of $\vec{B}_{\rm t}$ over the sky, which, however, is essential to match
the correlated patterns seen in the \Planck\ maps of the dust $p$ and $\psi$
\citep{planck2014-XIX}. 
Because of this, we use a different method. We model $\vec{B}_{\rm t}$ with
Gaussian realizations on the celestial sphere, computed for an angular power
spectrum $C_\ell$ scaling as a power-law $\ell^{\alpha_{\rm M}} $ for
$\ell \ge 2$.  The amplitude of the spectrum is parametrized 
by  the ratio $f_{\rm M}$ between the standard deviation of $|\vec{B}_{\rm t}|$
and $|\vec{B}_0|$.

Our spectrum does not have a low $\ell$ cut-off, which would represent 
the scale of energy injection of the turbulent energy cascade. Here, since we
compare the model and the data over a field with an angular
extent of $60^\circ$ (about 1\,radian), we implicitly assume that the
injection scale is larger than, or comparable to, the scale height of the dust
emission \citep[approximately 200\,pc,][]{Drimmel01}. 
The scale of the warm ionized medium (WIM) is larger \citep[about
1--1.5\,kpc,][]{Gaensler2008}, but the WIM is not a major component of the dust
emission from the diffuse ISM \citep{planck2013-XVII}. 
The range of distances involved in the modelling of dust polarization at high
Galactic latitudes is small, because there is little interstellar matter within
the local bubble, i.e., within 50--100\,pc of the Sun \citep{Lallement14}. 
The local bubble may extend to larger distances towards the Galactic poles,
but this possibility is not well constrained by existing data. In any case,
it is reasonable to assume that most of the dust emission at high Galactic
latitudes arises from a limited range of distances, which sets a rough
correspondence between angles and physical scales in our model.

To compute the Stokes parameters, we approximate the integration
along the line of sight (LOS) with a sum over a set of $N$ polarization
layers with independent realizations of $\vec{B}_{\rm t}$. 
The layers are a phenomenological means to 
represent the variation of $\vec{B}_{\rm t}$ along the LOS.  Our modelling of
$\vec{B}_{\rm t}$ is continuous over the celestial sphere, while we use a set
of independent orientations along the LOS. At first sight, this may be
considered as physically inconsistent. However, in Sect.~\ref{sec:discussion},
we relate the polarization layers to the density structure and to the correlation length of  $\vec{B}_{\rm t}$
along the LOS. Our modelling does
not take into account explicitly the density structure of matter along the LOS; 
the source function (presented in Eqs.~\ref{eq:simulated_Q} and
\ref{eq:simulated_U}) is assumed to be constant along the LOS. 
It also ignores the alignment observed between the filamentary structure of the diffuse ISM and the
magnetic field.

\subsection{Data fitting in three steps: A, B, and C}
\label{subsec:data_fit}

In the following two sections, we present three steps in our data-fitting,
labelled steps A, B, and C.  Step A only takes into account the mean field
$\vec{B}_0$. In Sect.~\ref{sec:ordfield}, we determine the orientation of
$\vec{B}_0$ by fitting the regular patterns seen in the
\Planck\ $Q_{353}$ and $U_{353}$ maps shown in Fig.~\ref{fig:data353}.
The other two models involve both $\vec{B}_0$ and $\vec{B}_{\rm t}$,
as required to reproduce the 1-point statistics of $\psi$ and $p$. 
In step B (Sect.~\ref{sec:turbfield}), $\vec{B}_{\rm t}$ is computed from
random realizations on the sphere.  In this model, the depolarization due to
changes in the orientation of $\vec{B}_{\rm t}$ along the LOS is accounted
for with an $F$ factor in Eq.~(\ref{eq:p_obs}) that is uniform over the sky.
This simplifying assumption is often made in analysing polarization data. 
Step C in Sect.~\ref{ssec:levels} is an extension of step B, where we
introduce variations of the $F$ factor over the sky
by summing Stokes parameters over $N$ polarization layers along the LOS. 

Our model has six parameters: the two coordinates defining the
orientation of $\vec{B}_0$; $f_{\rm M}$ quantifying the dispersion of
$\vec{B}$ around $\vec{B}_0$; the number of layers, $N$; the index
$\alpha_{\rm M}$; and the effective polarization fraction of dust emission,
$p_{0}$.  The parameters are not all fitted simultaneously because they
are connected to the data in different ways. The coordinates of $\vec{B}_0$
relate to the large-scale patterns in the $Q_{353}$ and $U_{353}$ maps
and they do not depend on the other parameters. The triad of parameters
$f_{\rm M}$, $N$, and $\alpha_{\rm M}$ describe statistical properties of the
polarization maps. We determine $f_{\rm M}$, $N$, and $p_0$ simultaneously
by fitting the 1-point statistics of both $\psi$ and $p$. To constrain
$\alpha_{\rm M}$ it is necessary to use 2-point statistics (i.e., power
spectra); this is not done in this paper, but will be the specific topic of a
future paper. 


\section{The mean orientation of the magnetic field}\label{sec:ordfield}

In this step A of our data modelling, we determine the orientation of the
mean field $\vec{B}_0$, ignoring $\vec{B}_{\rm t}$. 

\subsection{Description of step A}\label{ssec:modelA1}

\begin{figure}
\includegraphics[width=8.5cm]{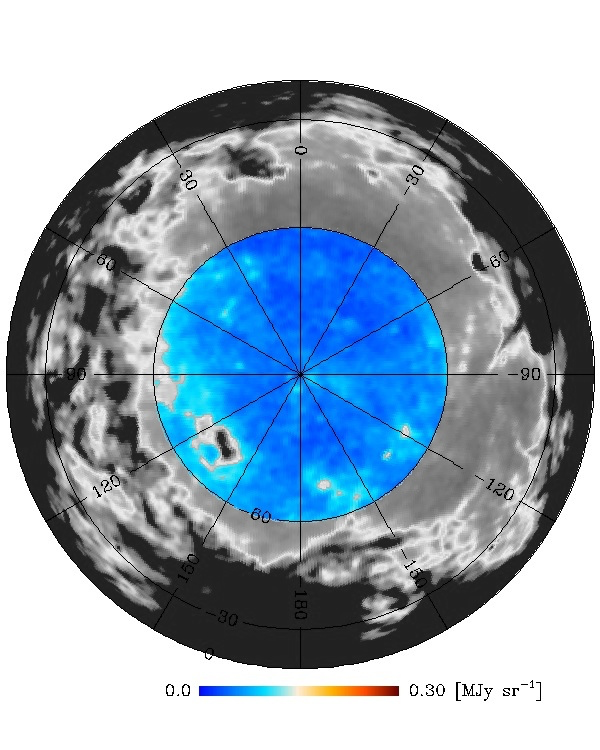}
\caption[]{Same as in the left panel of Fig.~\ref{fig:data353}, but
  now highlighting the $b < -60^\circ$ region, excluding the brightest clouds
  (in grey on the image) that has been used to fit step A to the data.}
\label{fig:intensity_mask}
\end{figure}


\begin{figure*}
\begin{tabular}{r l}
\includegraphics[width=8.5cm]{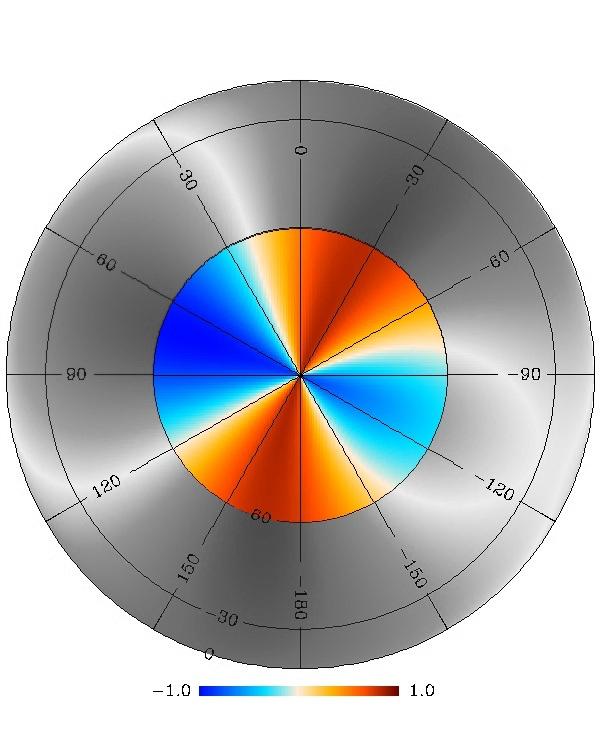}
& \includegraphics[width=8.5cm]{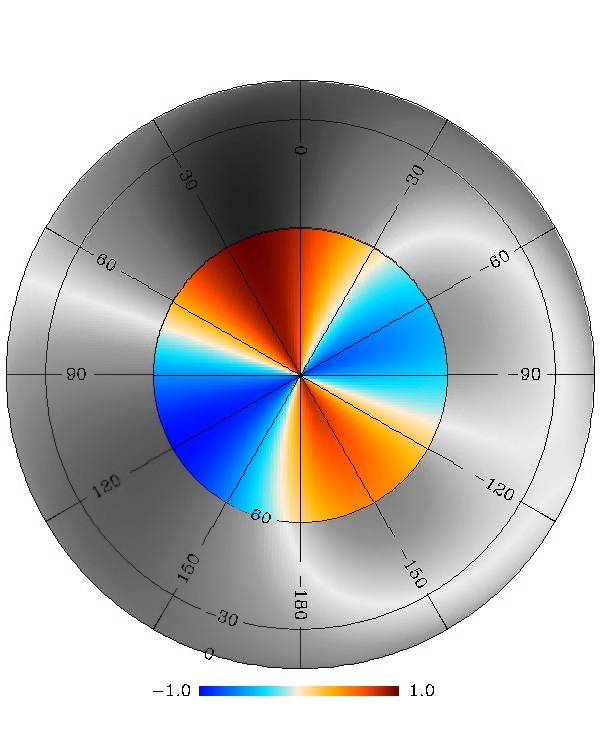}\\
\end{tabular}
\caption[]{Step A: orthographic projections of $q_{\rm A}$ (left) and
  $u_{\rm A}$ (right) centred on the south Galactic pole, for the best-fit
  direction of the uniform GMF towards $(l_0, b_0) = (70^\circ, 24^\circ)$.
  The sky at $b> -60^\circ$ is masked here.}
\label{fig:modelsuni}
\end{figure*}

We show that the ordered magnetic field produces
well-defined polarization patterns in the $Q_{353}$ and $U_{353}$ maps,
resulting from the variation across the observed region of the angle between
the LOS and the ordered field. 

Given a Cartesian reference frame {\it xyz},
each point on the sphere can be identified by a pair of angular
coordinates, hereafter the Galactic longitude and latitude,
$l$ and $b$. 
The reference frame is chosen to be centred at the observer with
$\vec{\hat{z}}=(0,0,1)$ pointing towards the
north Galactic pole, $\vec{\hat{x}}=(1,0,0)$ towards the Galactic
centre, and $\vec{\hat{y}}=(0,1,0)$ towards positive Galactic longitude.

We define the uniform direction of $\vec{B}_0$ through the unit vector
$\vec{\hat{B}}_{0}$, which depends on the pair of coordinates ($l_0$,$b_0$) as
$\vec{\hat{B}}_{0}=(\cos{l_0}\,\cos{b_0},\,\sin{l_0}\,\cos{b_0},\,\sin{b_0})$.
We define the generic LOS unit vector $\vec{\hat{r}}$ as
$(\cos{l}\,\cos{b},\,\sin{l}\,\cos{b},\,\sin{b})$ on a full-sky \healpix\ grid.

Combining $\vec{\hat{r}}$ and $\vec{\hat{B}}_{0}$, we can derive the POS 
component of $\vec{\hat{B}}_{0}$, $\vec{\hat{B}}_{0 \perp}$, as 

\begin{equation}\label{eq:bperp}
\vec{\hat{B}}_{0 \perp} = \vec{\hat{B}}_0 - \vec{\hat{B}}_{0 \parallel}
 = \vec{\hat{B}}_0 - (\vec{\hat{B}}_0 \cdot \vec{\hat{r}}) \vec{\hat{r}},
\end{equation} 
where $\vec{\hat{B}}_{0 \parallel}$ is the component of $\vec{\hat{B}}_0$
along $\vec{\hat{r}}$.  In order to define the $\psi$ and $\gamma$ angles for
a given $\vec{\hat{r}}$, we need to derive the north and east directions,
tangential to the sphere, which correspond to
\begin{align}
\label{eq:endirections}
 & \vec{\hat{n}}=\frac{(\vec{\hat{r}} \times \vec{\hat{z}}) \times
  \vec{\hat{r}}}{|(\vec{\hat{r}}
  \times \vec{\hat{z}}) \times \vec{\hat{r}}|}, \nonumber \\
 & \vec{\hat{e}} = \frac{-\vec{\hat{r}} \times \vec{\hat{n}}}{|\vec{\hat{r}}
  \times \vec{\hat{n}}|},
\end{align}
respectively. The polarization angle is the complement of that between
$\vec{\hat{B}}_{0\perp}$ and $\vec{\hat{n}}$, and $\gamma$ the angle between
$\vec{\hat{B}}_0$ and $\vec{\hat{B}}_{0\perp}$. From Eqs.~(\ref{eq:bperp})
and (\ref{eq:endirections}), we derive
\begin{align}
\label{eq:psigammam}
& \psi_{\rm A}=90^\circ - \arccos{\left(\frac{\vec{\hat{B}}_{0\perp} \cdot
   \vec{\hat{n}}}{|\vec{\hat{B}}_{0\perp}|}\right) }, \nonumber \\
& \cos^2{\gamma_{\rm A}}=1-|\vec{\hat{B}}_0 \cdot \vec{\hat{r}}|^2,
\end{align}
where the subscript ``A'' stands for step A, and the sign of $\arccos$
is imposed by the sign of $\vec{\hat{B}}_{0\perp} \cdot \vec{\hat{e}}$.

Using Eqs.~(\ref{eq:observables}) and (\ref{eq:p_obs}), we can produce an
analytical expressions for the modelled Stokes parameters normalized
to the total intensity times $p_0F$, 
$q_{\rm A}$ and $u_{\rm A}$, as follows:
\begin{align}
\label{eq:stokesA}
& q_{\rm A} = \cos^2{\gamma_{\rm A}}\cos{2\psi_{\rm A}}; \nonumber \\
& u_{\rm A} =- \cos^2{\gamma_{\rm A}}\sin{2\psi_{\rm A}}.
\end{align}
We stress that $q_{\rm A}$
and $u_{\rm A}$ only show patterns generated by projection
effects. For illustration, in Fig.~\ref{fig:Unimodel} we present maps
of $q_{\rm A}$ and $u_{\rm A}$ for a uniform direction of the GMF towards
$(l_0,b_0)=(80^\circ,0^\circ)$, roughly the direction inferred
from starlight polarization data \citep{Heiles96}.  We note that the total
intensity of dust emission also depends on the GMF geometry
\citep{planck2014-XX}. However, as detailed in Appendix~\ref{app:approx},
this is a small effect that does not alter our results. 

\subsection{Fitting step A to the \planck\ data}\label{ssec:modelA2}

At first glance, the ``butterfly'' patterns in the $Q_{353}$ and $U_{353}$
maps around the south Galactic pole in
Fig.~\ref{fig:data353} resemble those produced with step A
in Fig.~\ref{fig:Unimodel}. In order to find the orientation of
$\vec{\hat{B}}_0$ that best fits the data, we explore the space of Galactic
coordinates for ($l_0,b_0$), spanning Galactic longitudes between $0^\circ$
and $180^\circ$, and latitudes between $-90^\circ$ and $90^\circ$.
From Eqs.~(\ref{eq:observables}), (\ref{eq:p_obs}), and (\ref{eq:stokesA}),
we simultaneously fit step A to $\StokesQ_{353}$ and $\StokesU_{353}$ with the
corresponding errors, as
\begin{align}
\label{eq:stokesAfit}
& Q_{353}= p_{0{\rm,A}}q_{\rm A}D_{353},\nonumber \\
& U_{353}= p_{0{\rm,A}}u_{\rm A}D_{353},
\end{align}
where the factor $p_{0{\rm,A}}$ represents an average of the product
$p_0F$ in Eq.~(\ref{eq:p_obs}) over the region where we perform the fit.  
For each $(l_0,b_0)$ pair we perform a linear fit to determine
$p_{0{\rm,A}}$.  The fit is carried out for the southern polar cap at
$b<-60^\circ$, after masking the most intense localized structures around
the south Galactic pole, as shown in Fig.~\ref{fig:intensity_mask}. To remove
these regions from the analysis, we fit a Gaussian profile to the histogram
of pixel values of $D_{353}$ below $b=-60^\circ$. 
We then mask all pixels with $D_{353}> \bar{x}_D + 4\,\sigma_{\bar{x}_D}$, where
$\bar{x}_D$ and $\sigma_{\bar{x}_D}$ are the mean and the
standard deviation of the Gaussian fit.

\begin{figure*}
\begin{tabular}{r l}
\includegraphics[width=8.5cm]{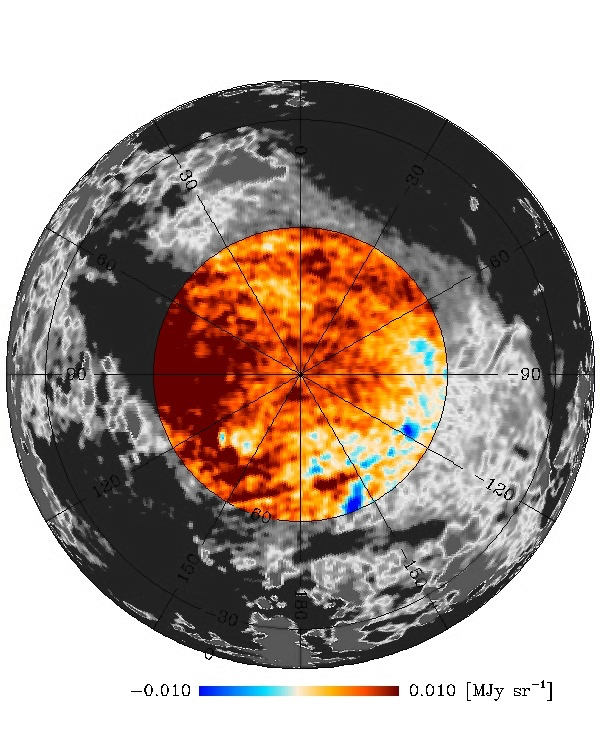}
 & \includegraphics[width=8.5cm]{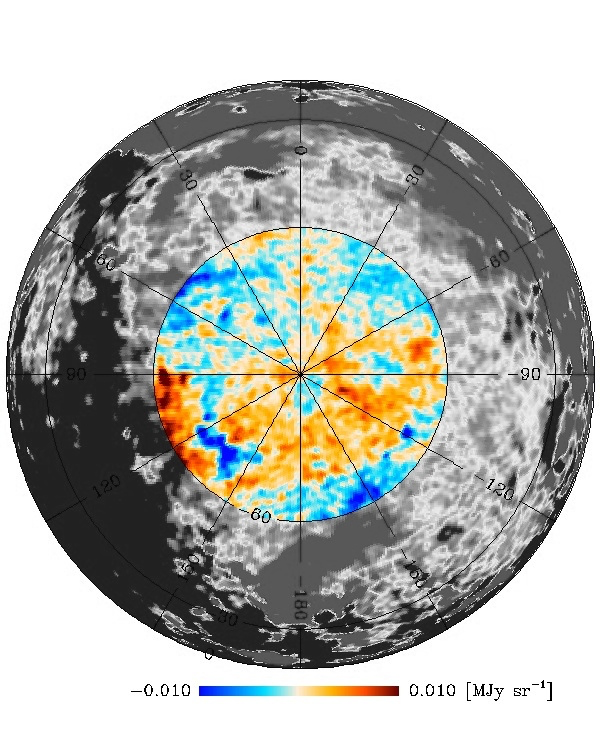}\\
\end{tabular}
\caption[]{Orthographic projections centred on the south Galactic
  pole of $\StokesQ^{R}_{353}$ ({\it left}) and $\StokesU^{R}_{353}$
  ({\it right}), the Stokes parameters in a reference frame rotated with
  respect to the best-fit direction of the uniform component of the GMF
  towards $(l_0,b_0)=(70^\circ,24^\circ)$. The sky for the masked $b>-60^\circ$
  region appears in grey.}
\label{fig:dataROT353}
\end{figure*}

The fit is done over an area of $2652\,$deg$^2$, corresponding to 2652 independent data beams. 
Since the number of parameters is 3, the number of degrees of freedom, $N_{\rm dof}$, is large. 
We find a best-fit direction of the mean GMF towards Galactic coordinates
$l_0=70^\circ\pm 5^\circ$ and $b_0=24^\circ \pm 5^\circ$.  The value of $p_{0{\rm,A}}$ corresponding to this direction is
$(12 \pm 1)\%$. The statistical errors are small but there
are significant uncertainties on the three parameters from residual, uncorrected,
systematic effects in the data. We quote these  uncertainties, which we estimated
repeating the fit on maps produced with ten different subsets of the data
\citep{planck2014-ES}.  We notice that,
because of the 180\deg\ ambiguity in the definition of $\psi$, the opposite
direction $(l_0 + \pi, -b_0)$ is an equivalent solution of our fit.
However, the chosen solution is the closest to the mean GMF direction derived
from observations of pulsars in the solar neighbourhood
\citep{Rand89,Ferriere15}, which, unlike dust polarization are sensitive to
the sign of the GMF.  Our determination of $l_0$ is in agreement with earlier values derived from
starlight polarization \citep[e.g.][]{Heiles96}. The positive value of $b_0$ is consistent with the positive sign of the median value of 
rotation measures derived from observations of extragalactic radio sources in the direction of the southern Galactic cap \citep{Taylor09,Mao10}. 
For illustration, we show the best-fit model maps of
$q_{\rm A}$ and $u_{\rm A}$ around the south pole in Fig.~\ref{fig:modelsuni}. 

We note that the obtained value of $p_{0{\rm,A}}$ is a substantial fraction
of the maximum $p$ ($> 18\,\%$) reported in \citet{planck2014-XIX} at
intermediate Galactic latitudes. This result confirms that dust
polarization is important at high
Galactic latitudes. We also stress that this value of $p_{0{\rm,A}}$ is 
only a lower limit to the effective dust polarization fraction, because step A
does not take into account any depolarizing effects along the LOS, associated
with variations of the GMF orientation.


\section{The turbulent component of the magnetic field}\label{sec:turbfield}

The \planck\ maps show structures in polarization on a wide range of scales
(Fig.\ref{fig:data353}), not accounted for by the single field orientation
of step A, which we associate with the turbulent component of the magnetic field  $\vec{B}_{\rm t}$. In
Sects.~\ref{ssec:disp_psi} and \ref{ssec:pUNI}, $\vec{B}_{\rm t}$ is
assumed to vary only across the sky (step B),
while in Sect.~\ref{ssec:levels}, we take into account its variations 
both across the sky and along the LOS (step C).

\subsection{Step B: dispersion of the polarization angle}\label{ssec:disp_psi}

\begin{figure}
\includegraphics[width=9.5cm]{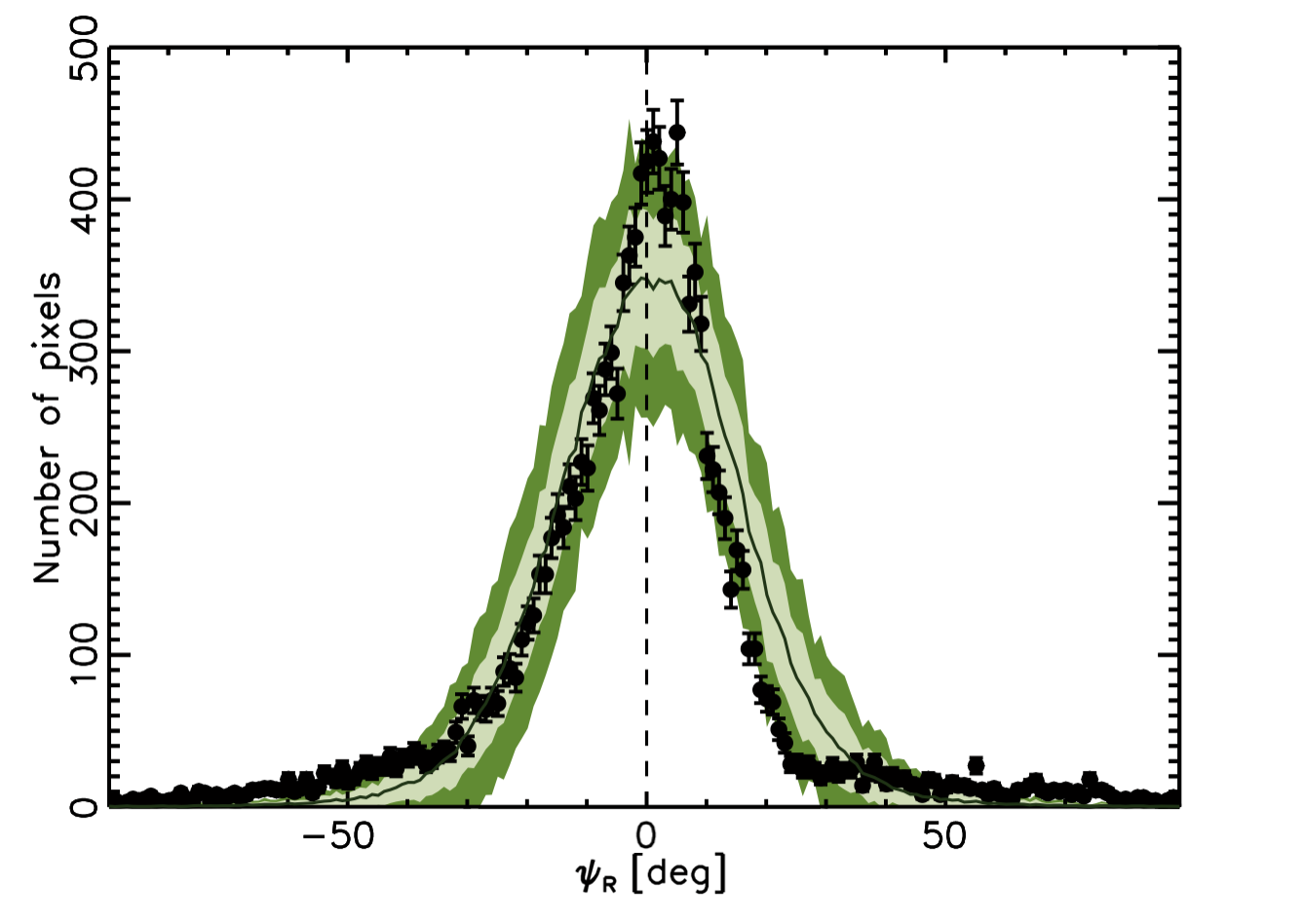}\\
\includegraphics[width=9.5cm]{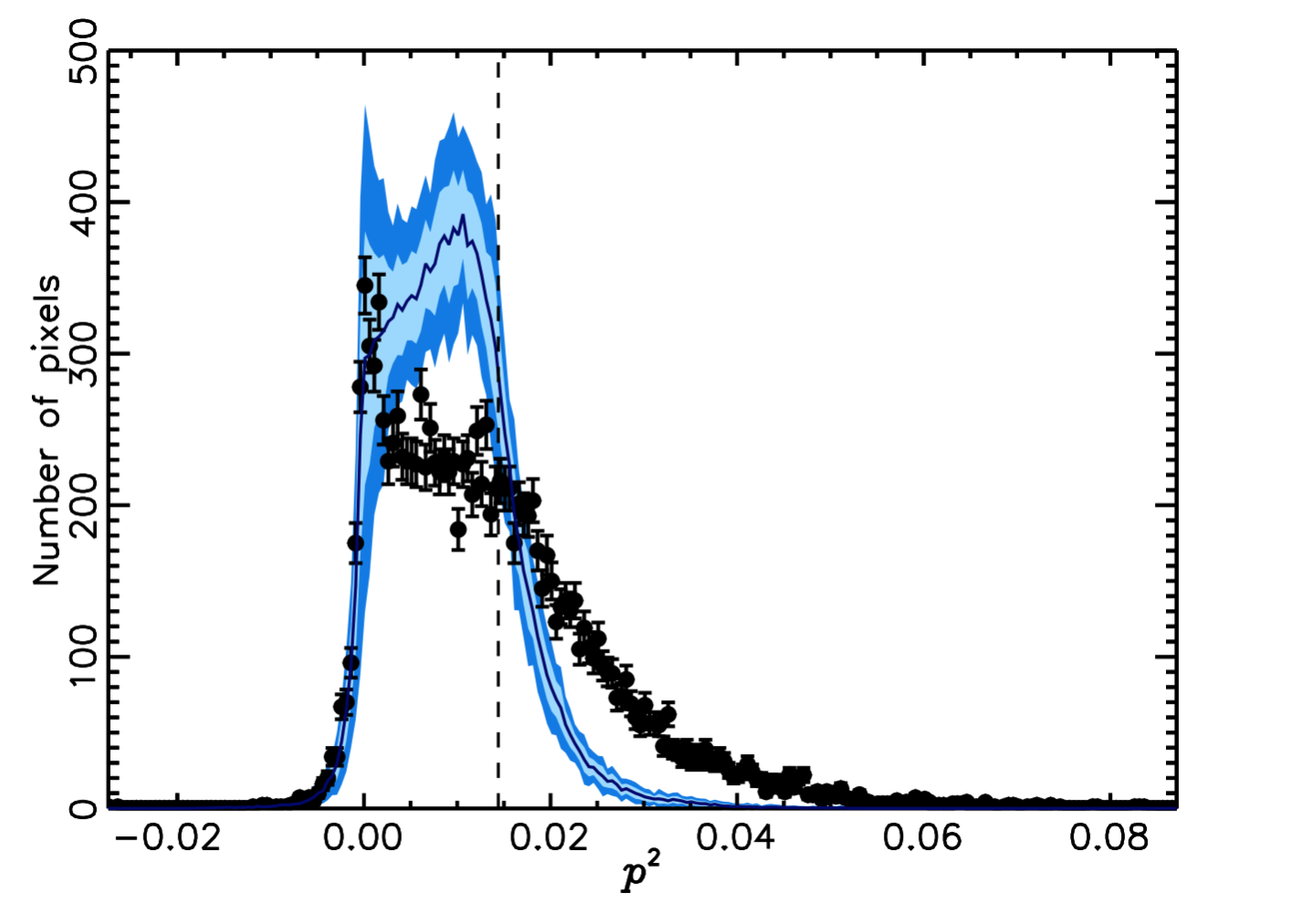}\\ 
\caption[]{Results of step B. {\it Top}: Histogram of $\psi_{\rm R}$, the
  polarization angle inferred from the Stokes parameters
  rotated with respect to the best-fit uniform direction of the
  GMF ($\StokesQ^{R}_{353}$ and $\StokesU^{R}_{353}$), over the southern
  Galactic cap (black dots). The error bars represent the Poisson
  noise within each bin of the histogram. The green line represents the mean
  of the step B results for $f_{\rm M}=0.4$ over $20$ different realizations. The green
  shaded regions correspond to $\pm1\,\sigma$ (light green) and
  $\pm2\,\sigma$ (dark green) variations of the model.
  {\it Bottom}: Histogram of $p^2$ obtained when combining the Year~1 and
  Year~2 maps (black dots). The error bars here represent the Poisson
  noise within each bin of the histogram. Step B is now shown in blue.
  The dashed vertical line corresponds to a value of the polarization
  fraction of $12\,\%$.}
\label{fig:p2psi_uni}
\end{figure}

In Sect.~\ref{ssec:modelA2}, we found that the best-fit orientation of
$\vec{B}_0$ in step A is given by $(l_0,b_0)=(70^\circ,24^\circ)$. We
can now obtain maps of the corresponding normalized Stokes parameters,
$u_{0 \rm A}$ and $q_{0 \rm A}$, as well as a map of the associated
polarization angle
\begin{equation}\label{eq:psiA} 
\psi_{0\rm A}=\frac{1}{2}{\rm tan}^{-1}\left(-u_{0 \rm A},q_{0 \rm A}\right).
\end{equation}
The angle $\psi_{0\rm A}$ allows us to rotate, at each point on the sky,
the reference direction used to compute the Stokes parameters
($\StokesQ_{353}$,$\StokesU_{353}$). With this new reference, the 
$q_{\rm A}$  map in Fig.~\ref{fig:modelsuni} would be that of $\cos^2 \gamma_{\rm A}$,
and $u_{\rm A}$ would be null (see Eq.~\ref{eq:stokesA}). 
To obtain the rotated values
$\StokesQ^{R}_{353}$ and $\StokesU^{R}_{353}$, we apply to the data
the following rotation matrix \citep[e.g.,][]{delabrouille2009}:
\begin{equation}\label{eq:rotpsi}
\left(\begin{array}{c}
Q^{\rm R}_{353}\\
U^{\rm R}_{353} \end{array}\right)=\left(\begin{array}{ccc}
\cos{2 \psi_{0\rm A}} & \sin{2 \psi_{0\rm A}} \\
-\sin{2 \psi_{0\rm A}} & \cos{2 \psi_{0\rm A}}\end{array}\right)
 \left(\begin{array}{c}
Q_{353}\\
U_{353} \end{array}\right).
\end{equation}

The maps of $\StokesQ^{R}_{353}$ and $\StokesU^{R}_{353}$ are shown
in Fig.~\ref{fig:dataROT353}, where the butterfly patterns, caused by
the uniform component of the GMF, are now removed by the change of reference.
The polarization angle that can be derived from $\StokesQ^{R}_{353}$ and
$\StokesU^{R}_{353}$ as
\begin{equation}\label{eq:psirot} 
\psi_{\rm R}=\frac{1}{2}{\rm tan}^{-1}(-U^{\rm R}_{353},Q^{\rm R}_{353} ),
\end{equation}
represents the dispersion of $\vec{B}_\perp$ around
$\vec{B}_{0\perp}$. The histogram of $\psi_{\rm R}$ for
$b<-60^{\circ}$, shown in the top panel of Fig.~\ref{fig:p2psi_uni}
(black dots with Poisson noise as error bars), has a $1\,\sigma$
dispersion of 12\deg.

To characterize $\vec{B}_{\rm t}$, it is necessary to account for projection
effects \citep{Falceta08,planck2014-XXXII}. 
\citet{planck2014-XXXII} describes a geometric
model, which we use in this paper to
characterize the 3D dispersion of $\vec{B}$ with respect to
$\vec{B}_{0}$, given the histogram of $\psi_{\rm R}$. 
Each component of $\vec{B}_{\rm t}$ is obtained with an independent realization of a
Gaussian field with an angular power spectrum equal to a power law of index
$\alpha_{\rm M}$, for multipoles $\ell \ge 2$. 
The degree of alignment between $\vec{B}$ and $\vec{B}_{0}$ is
parameterized by $f_{\rm M}$, which represents the ratio between
the strengths of the turbulent and mean components of the GMF.
 
In the top panel of Fig.~\ref{fig:p2psi_uni}, we show that
for $f_{\rm M}=0.4$ the model reproduces the histogram of $\psi_{\rm R}$
fairly well.  We computed $20$ different Gaussian
realizations to take into account the statistical variance of the model. 
The green line represents the average of the $20$ realizations, whereas the
green shaded regions are the $\pm1\,\sigma$ (light) and $\pm2\,\sigma$ (dark)
variations of the model.  In these calculations, as in
\citet{planck2014-XXXII}, the spectral index $\alpha_{\rm M}$ has a value of
$-1.5$.  This specific choice does not impact the distribution of
$\psi_{\rm R}$, or that of $p$. However, we note that the variance of the histogram,
i.e., the dispersion of histogram values between independent realizations, 
increases for decreasing values of $\alpha_{\rm M}$. 
\begin{figure}
\includegraphics[width=9.5cm]{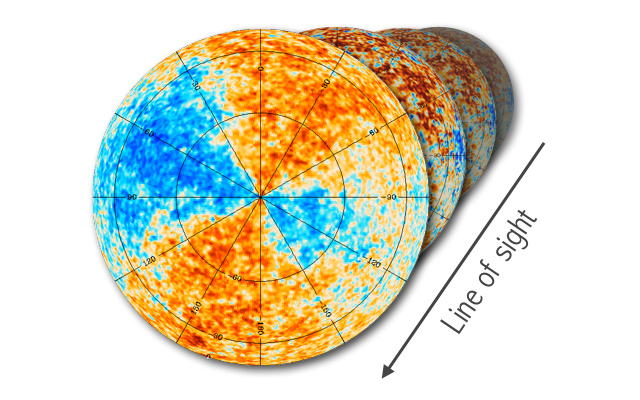}\\
\caption[]{Cartoon illustrating, for step C, the integration
  of $q_{\rm C}$ along the LOS, with four distinct
  polarization layers for the same value of $f_{\rm M}$ and the 
  same mean orientation of the GMF. Each map in this cartoon is a realization of the model.}
\label{fig:layers}
\end{figure}
\subsection{Step B: histogram of the polarization fraction}\label{ssec:pUNI}

We showed that the structure of the GMF on the sphere allows us to reproduce
$\psi_{\rm R}$ over the southern Galactic cap. Here, we characterize
the distribution of $p$ at $b< -60^\circ$ and we show that step B is
not sufficient to describe the data. 

As already discussed above, the noise bias on $p$ represents an
intrinsic problem. To circumvent 
it, we compute unbiased values of $p^2$ by
multiplying Stokes parameters from subsets of the data. Doing this, instead of using 
$p_{\rm MAS}$ as in Sect.~\ref{ssec:poldata},  gives us
control over the level of noise in the data, as we now
demonstrate.  We use the year-maps (denoted by the
indices ``Y1'' and ``Y2''), which have uncorrelated instrumental noise, and
compute $p^2$ as
\begin{equation}\label{eq:p2d} 
p^2=\frac{Q^{\rm Y1}_{353}Q^{\rm Y2}_{353}
 +U^{\rm Y1}_{353}U^{\rm Y2}_{353}}{(D_{353})^2}.
\end{equation}
We also estimate $p^2$ from the so-called ``DetSet'' maps \citep[made from different subsets of detectors, see][]{planck2014-ES}, and we
find good agreement between the two estimates using distinct subsets of the
data. 
\begin{figure*}
\center
\includegraphics[width=18cm]{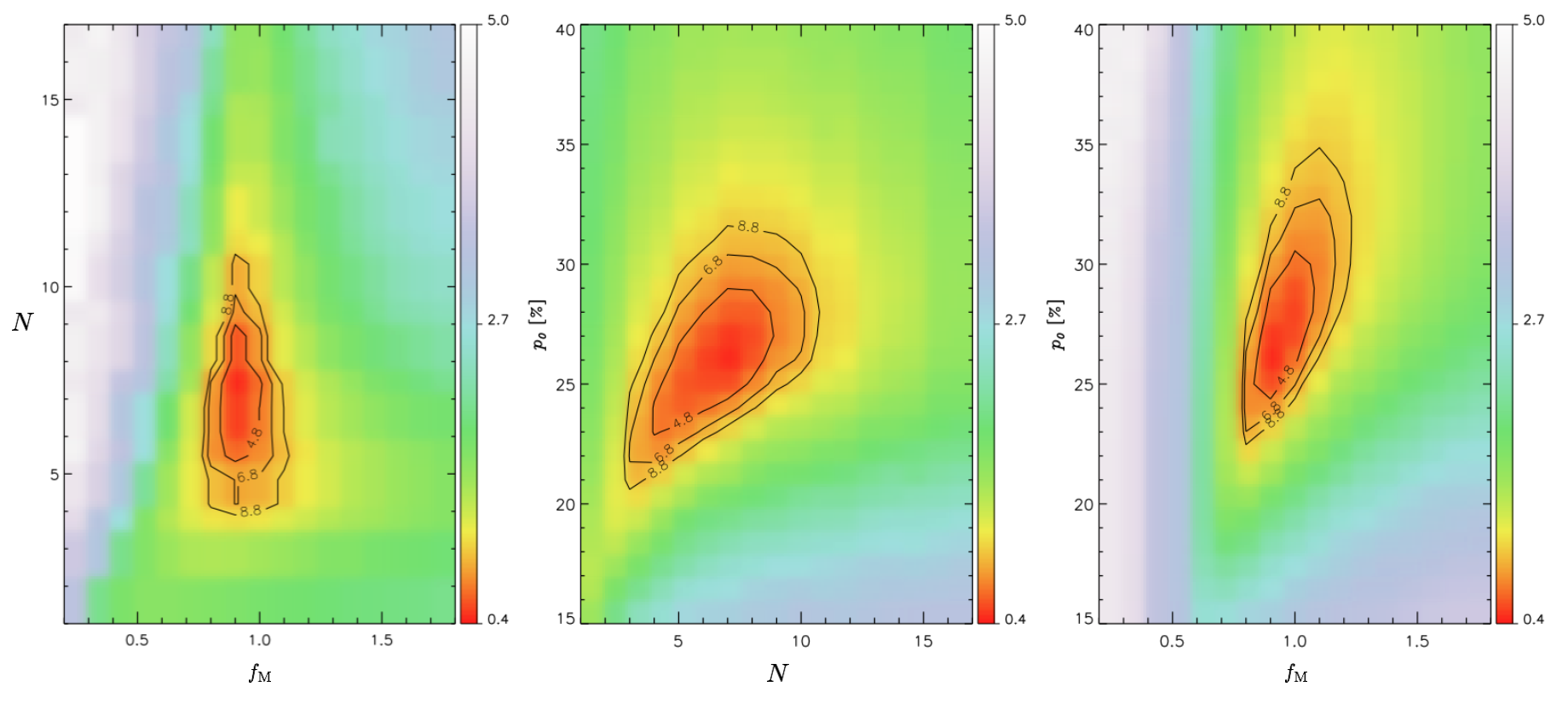}\\
\caption[] {Results of step C. Maps of $\chi^2_{\rm tot}$ from the fit of
  step C to the data for $p_0=26\,\%$ ({\it left}), for $N=7$ ({\it centre}),
  and for $f_{\rm M}=0.9$ ({\it right}). The three maps show in
  colours and with contours the quantity ${\rm log}_{10} (\chi^2_{\rm tot})$.}
\label{fig:chi2C}
\end{figure*}
In order to model $p^2$, we make use of the results obtained from
fitting steps A and B to the data. Given $\vec{B}_0$, pointing towards
$(l_0,b_0)=(70^\circ,24^\circ)$, we add $\vec{B}_{\rm t}$ to it with
normalization parameter $f_{\rm M}=0.4$. In doing so, we now produce the two
variables $q_{\rm B}$ and
$u_{\rm B}$, as $q_{\rm A}$ and
$u_{\rm A}$ in Eq.~(\ref{eq:stokesA}), where now the angles take into
account the turbulent component of the GMF. 
We then make realizations of the \planck\ statistical
noise ($n_{Qi}$ and $n_{Ui}$, with $i=1,2$), and, as in
Eq.~(\ref{eq:stokesAfit}), we produce two pairs of independent
samples of modelled Stokes $Q$ and $U$ as
\begin{align}
\label{eq:stokesnoise}
& Q_{{\rm M}i}= p_0q_{\rm B}D_{353} + n_{Qi},\nonumber \\
& U_{{\rm M}i}= p_0u_{\rm B}D_{353} + n_{Ui},
\end{align}
in which $i=1,2$ and $p_0=12\,\%$. Thus, the
modelled $p^2$ results from
\begin{equation}\label{eq:p2m} 
p_{\rm M}^2=\frac{Q_{\rm M1}Q_{\rm M2}+U_{\rm M1}U_{\rm M2}}{(D_{353})^2}.
\end{equation}
In the bottom panel of Fig.~\ref{fig:p2psi_uni}, we show the comparison
between the histograms of $p^2$ for the data (black dots) and for the
model. In particular, we present the average over 20
realizations of step B (blue line) and the corresponding
$\pm1\,\sigma$ (light blue shaded region) and $\pm2\,\sigma$ (dark blue)
variations. The dashed vertical line refers to the value of $p_0=12\,\%$.   
We notice that our modelling of $p^2$ seems to appropriately take into account 
the data noise, since it nicely fits the negative
$p^2$ values, which result from noise in the combination of
the individual year maps.

However, from Fig.~\ref{fig:p2psi_uni} it is clear that our description of the
GMF structure using step B does not provide a satisfactory
characterization of the distribution of $p^2$. The data show a more
prominent peak in the distribution towards very low $p^2$ values than
seen in the model, for which the histogram peaks near the value of $p_0$.
Moreover, the large dispersion in the data, also found by
\citet{planck2014-XIX} at intermediate Galactic
latitudes, produces a long tail in the distribution towards high
values of $p^2$, which is not reproduced by the model.

\subsection{Step C: line-of-sight depolarization}\label{ssec:levels}

Now we consider the effect of depolarization, associated with variations of
the GMF orientation along the LOS. This additional step is essential to
account for the dispersion of $p$ and correctly estimate the amplitude of
the turbulent component of the GMF with respect to its
mean component, because the dispersion of the polarization angle is reduced
by averaging along the LOS \citep{Myers91,Jones92,Houde09}.

\begin{figure}
\includegraphics[width=9.5cm]{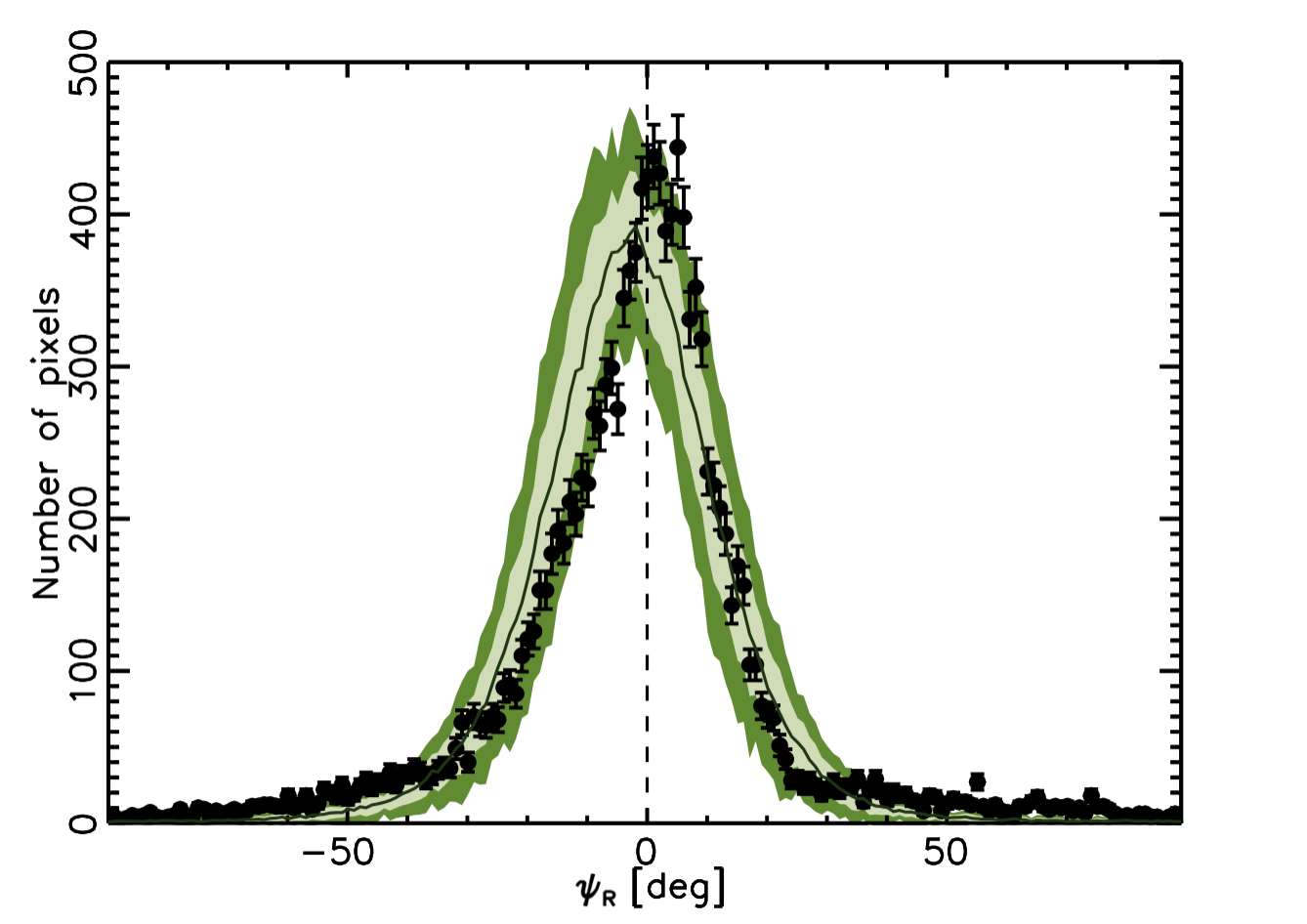}\\
\includegraphics[width=9.5cm]{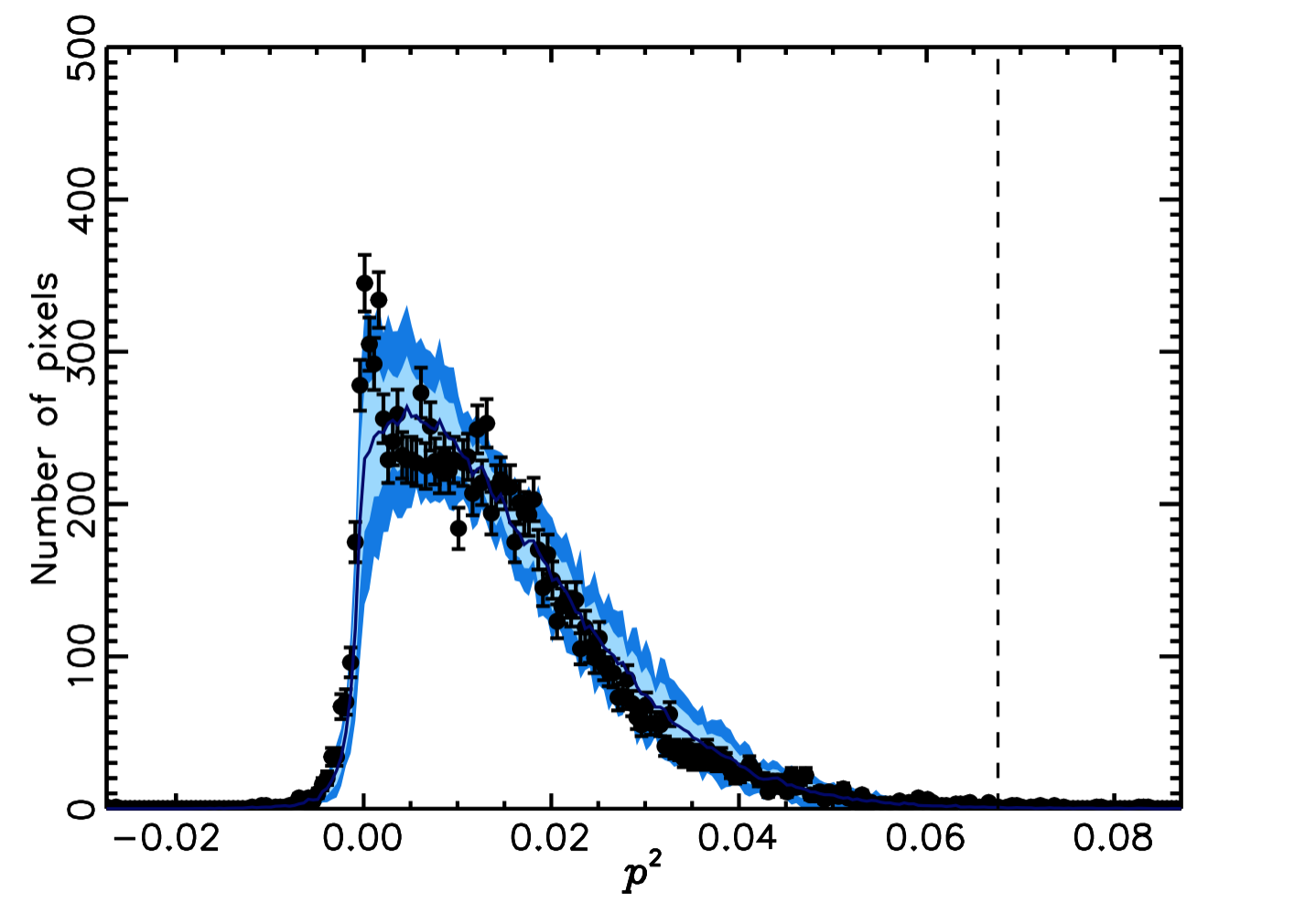}\\ 
\caption[]{Results of step C. This is the same as in Fig.~\ref{fig:p2psi_uni},
  but with the model histograms now corresponding to step C with
  $f_{\rm M}=0.9$, $N=7$, and $p_0=26\,\%$ (dashed-vertical line).}
\label{fig:p2psi_layers}
\end{figure}

Figure~\ref{fig:layers} illustrates step C
with a simple cartoon. In order to account for the LOS integration
that characterizes the polarization data, we produce $N$ distinct
maps of $q_{{\rm B},i}$ and $u_{{\rm B},i}$ (with $i$ from 1 to $N$),
for a common, but freely varying value of $f_{\rm M}$, while fixing
$\alpha_{\rm M}= -1.5$ (as in step B), and for the best-fit orientation of
$\vec{B}_0$ obtained with step A. The Gaussian realizations of
$\vec{B}_{\rm t}$ are different for each layer. All layers have the same
$\vec{B}_0$ but an independent $\vec{B}_{\rm t}$ in Eq.~(\ref{eq:Bfield}).
Then, we model the LOS depolarization by averaging the Stokes parameters
over the $N$ layers as follows:
\begin{align}
\label{eq:intQU}
& q_{\rm C}= \frac{\sum^N_{i=1} q_{{\rm B},i}}{N};\nonumber \\
& u_{\rm C}= \frac{\sum^N_{i=1} u_{{\rm B},i}}{N}.
\end{align}
We follow the same procedure as in Sects.~\ref{ssec:disp_psi} and
\ref{ssec:pUNI}, with $q_{\rm B}$ and $u_{\rm B}$ replaced by $q_{\rm C}$ and
$u_{\rm C}$, to obtain model distributions of $p^2$ and $\psi_{\rm R}$. 

Given $\alpha_{\rm M}$, the modelled distributions of $p^2$ and $\psi_{\rm R}$
depend on three main parameters, namely $p_0$, $f_{\rm M}$, and $N$. 
We fit the data exploring the parameter spaces of
$p_0$ between $15\,\%$ and $40\,\%$ with steps of $1\,\%$, of $f_{\rm M}$
between $0.2$ and $1.8$ with steps of $0.1$, and of $N$ between $1$ and $17$
with steps of $1$.  
The distributions of $p^2$ and $\psi_{\rm R}$ have about 200  bins each. 
For each triad of parameters we compute maps of the
reduced $\chi^2$ for the combined $p^2$ and $\psi_{\rm R}$ fit, using
\begin{equation}\label{eq:chi2tot}
\chi^2_{\rm tot}=\chi^2_{p^2}+\chi^2_{\psi_{\rm R}},
\end{equation}
where in computing the $\chi^2$ distributions we fit the data with the
mean of the $20$ realizations, and we add their dispersion in quadrature
to the error bar of the observations. 
Fitting the distribution of $\psi_{\rm R}$ between $-40^\circ$ and
$40^\circ$ (where most of the data lie), we obtain a best fit for a minimum
$\chi^2_{\rm tot}$ of $2.8$,
for $p_0=26\,\%$, $f_{\rm M}=0.9$, and $N=7$.
In Fig.~\ref{fig:chi2C} we show three maps of $\chi^2_{\rm tot}$; each one
corresponds to the parameter space for two parameters given the best-fit value
of the third one. The $\chi^2_{\rm tot}$ maps reveal some
correlation among the three parameters. The variance of each model
among the $20$ different realizations represents the dominant
uncertainty of the fit, and it is correlated between the bins of the
histogram. Repeating the $\chi^2-$minimization for each one of the $20$
realizations, the fit constrains the range of values for the main
parameters to $0.8<f_{\rm M}<1$, $5<N<9$, and $23\,\%<p_{0}<29\,\%$.
Step C generates a mean value of the depolarization factor $F$ that is about 0.5, 
and thus leads to an estimate of $p_0$ twice larger than in step A.
The best-fit value of $(26\pm 3)\,\%$ is comparable with the maximum value
of the observed reported in \citet{planck2014-XIX}.

As in Fig.~\ref{fig:p2psi_uni}, the histograms of $p^2$ and $\psi_{\rm R}$
for the best-fit triad are shown in the bottom and top panels of
Fig.~\ref{fig:p2psi_layers}, respectively.
The top panel of Fig.~\ref{fig:p2psi_layers} shows that if we consider a few
($N\simeq7$) independent polarization layers along the
LOS, this provides us with an estimate of $f_{\rm M}$ that is closer to 
equality between the turbulent and mean components of the GMF than for
step B (for which $N=1$, see Sect.~\ref{ssec:disp_psi}). A value of
$f_{\rm M}=0.9$ with $N=1$ would generate a much broader distribution of
$\psi_{\rm R}$ than the observed one. 
The bottom panel of Fig.~\ref{fig:p2psi_layers} shows that
step C, unlike step B, can reproduce the histogram of $p^2$ quite well. The
combination of a small number of independent polarization layers
along the LOS produces the large dispersion in $p^2$ that is observed in
the data. 

\begin{figure}
\hspace{-0.2cm}
\includegraphics[width=9.8cm]{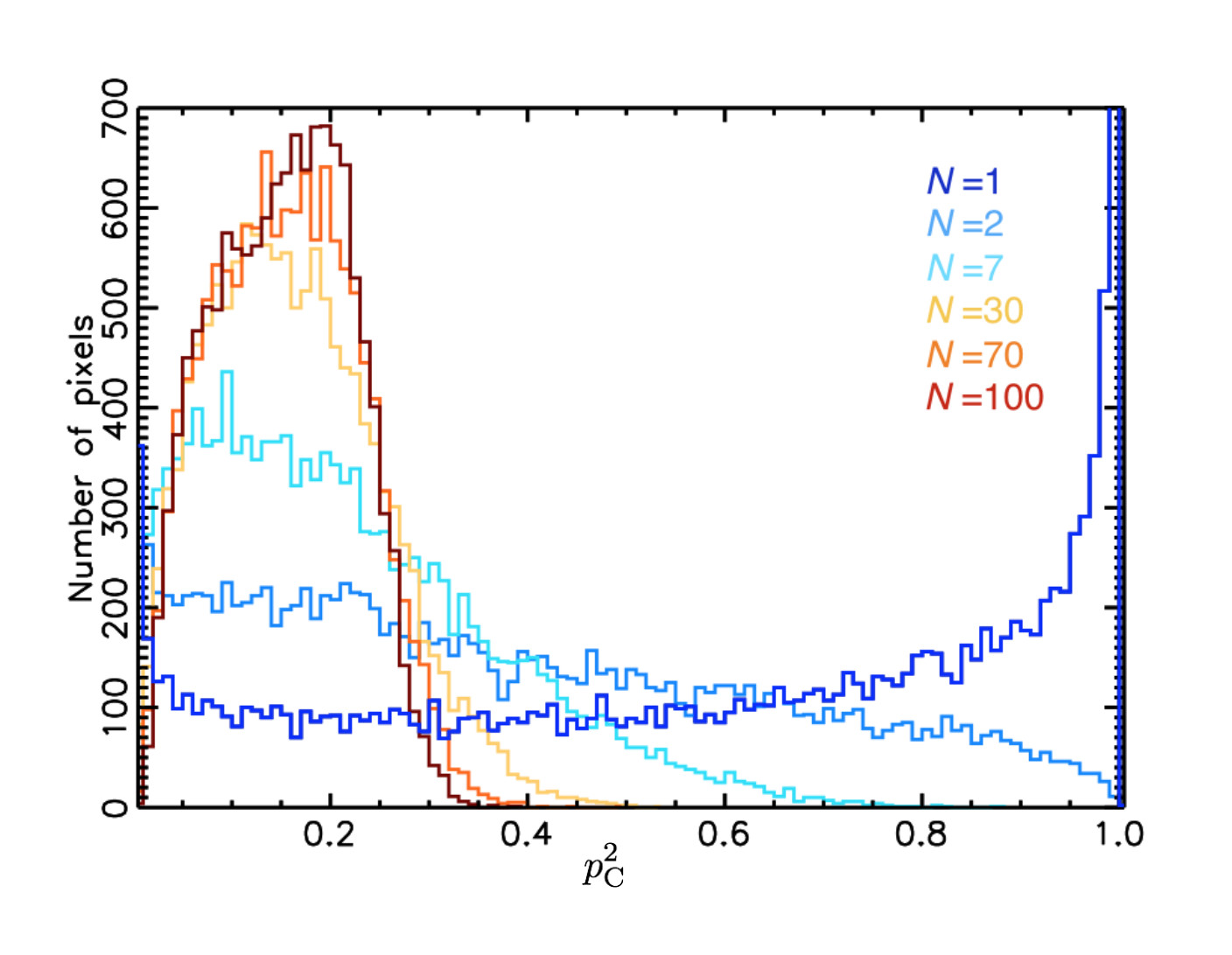}\\
\caption[]{Model histograms of $p^2_{\rm C}$ obtained around the south
  Galactic pole from step C, using $f_{\rm M}=0.9$, and with the value of $N$
  varying from 1 (dark blue) to 100 (dark red).}
\label{fig:Modelayers}
\end{figure}

Our results show that, in order to reproduce the $p^2$ distribution seen in
the data, only a small number of polarization
layers is needed. In Fig.~\ref{fig:Modelayers}, we present the effect
of changing $N$ on the distribution of $p^2$ obtained with step C as
$p^2_{\rm C}=q^2_{\rm C}+u^2_{\rm C}$. In this case noise is not added and
we fix $l_0=70^\circ$, $b_0=24^\circ$, and $f_{\rm M}=0.9$, but vary $N$
from $1$ (dark blue line) to $100$ (dark red line). The figure shows that for
an increasing number of layers, because of the central-limit theorem, the
model distributions tend to rapidly converge towards a low $p^2$
value, without the broad dispersion observed in the data. For large
values of $N$, the width of the $p^2$ distribution is dominated by
the projection factor, $\cos^2 \gamma$, in Eq.~(\ref{eq:stokesA}).
Note that the histogram of $p^2_{\rm C}$
for $N=1$ is not directly comparable with the modelled $p^2$ distribution
in Fig.~\ref{fig:p2psi_uni}, because it does not include noise.


\section{Discussion}\label{sec:discussion}

We have presented a phenomenological model that is able
to describe the $1-$point statistics of $p$ and $\psi$ for the \planck\ dust
polarization data around the south Galactic pole, using
a few parameters to describe the uniform and turbulent components of the
GMF. We stress that our model is not entirely physical and certainly not
unique. We made several assumptions, including: a single orientation of the
mean field $\vec{B}_0$; a uniform ratio $f_{\rm M}$ of the turbulent to
mean strengths of the GMF along the LOS; a fixed value for the number of
polarization layers, $N$, independent of the total dust intensity (unlike
what was considered by \citet{Jones92}); and isotropy of the turbulent
component, $\vec{B}_{\rm t}$. 
These assumptions restrict us from fitting the data over a larger portion of the sky than the southern Galactic cap.
For the time being, we limit our study to this sky area.
We now discuss the interpretation of our model in relation to the ISM physics
and we present future perspectives on the modelling.

\subsection{The density structure of the
 ISM}\label{ssec:disc_structures}

Our description of the turbulent component of the GMF along the LOS is based
on a finite number of independent layers, rather than on a continuous
variation computed from the power spectrum of the GMF, as was included
in some earlier models \citep[e.g.,][]{MAMD2008,ODea12,Planck2016-XLII}.
The density structure of the diffuse ISM provides one argument in favour of
this approximation.

If we are in practice observing a finite number of localized density structures
from the cold neutral medium (CNM) along the LOS, then the discretization of
the GMF orientation is appropriate. Such structures appear as extended features
on the sky in dust emission maps, with a power-law power spectrum. 
This statement is exemplified by the images and the power spectrum analysis
of the dust emission from the Polaris cloud in \citet{Mamd10}.
The superposition of such clouds fits with our model, where the angular
correlation is described with a continuous power spectrum, different from our 
ansatz for the radial correlation.
 
As shown for the diffuse ISM
\citep{Clark13,planck2014-XXXII,planck2015-XXXVIII,Kalberla16}, the GMF
orientation is correlated with the structure of matter as traced by \hi\ or
dust emission. Our modelling does not include the density structure of the ISM,
nor does it include the correlation between matter and the magnetic field
orientation; however, the polarization layers could phenomenologically
represent distinct matter structures along the
LOS.  In this interpretation the GMF orientations are not
completely uncorrelated. 
Although each CNM structure has a different turbulent component
of the GMF, they share the same mean component. This correlation
between the values of $\psi$ of individual structures and those measured for
the background emission in their surroundings is in fact observed in the
\planck\ data \citep{planck2014-XXXII}.

Observations of \hi\ in absorption and emission have shown that, in the solar
neighbourhood, about $60\,\%$ of all \hi\ arises from the warm neutral medium
(WNM) and gas that is out of thermal equilibrium \citep{Heiles2003}. 
Moreover, the diffuse ISM also includes the WIM, which accounts for about
$25\,\%$ of the gas column density \citep{Reynolds89}.
These diffuse and warm components of the ISM are expected to contribute to the
dust emission observed at high Galactic latitudes, both in intensity and
in polarization.  This contribution, which may be dominant, cannot be
described by a small number of localized structures. For such
media, the layers acquire a physical meaning if their spacing corresponds
approximately to the correlation length of the turbulent component of the GMF. 

\subsection{The correlation length of the magnetic field}\label{ssec:disc_corr}

\begin{figure}
\begin{tabular}{r l}
\includegraphics[width=9.5cm]{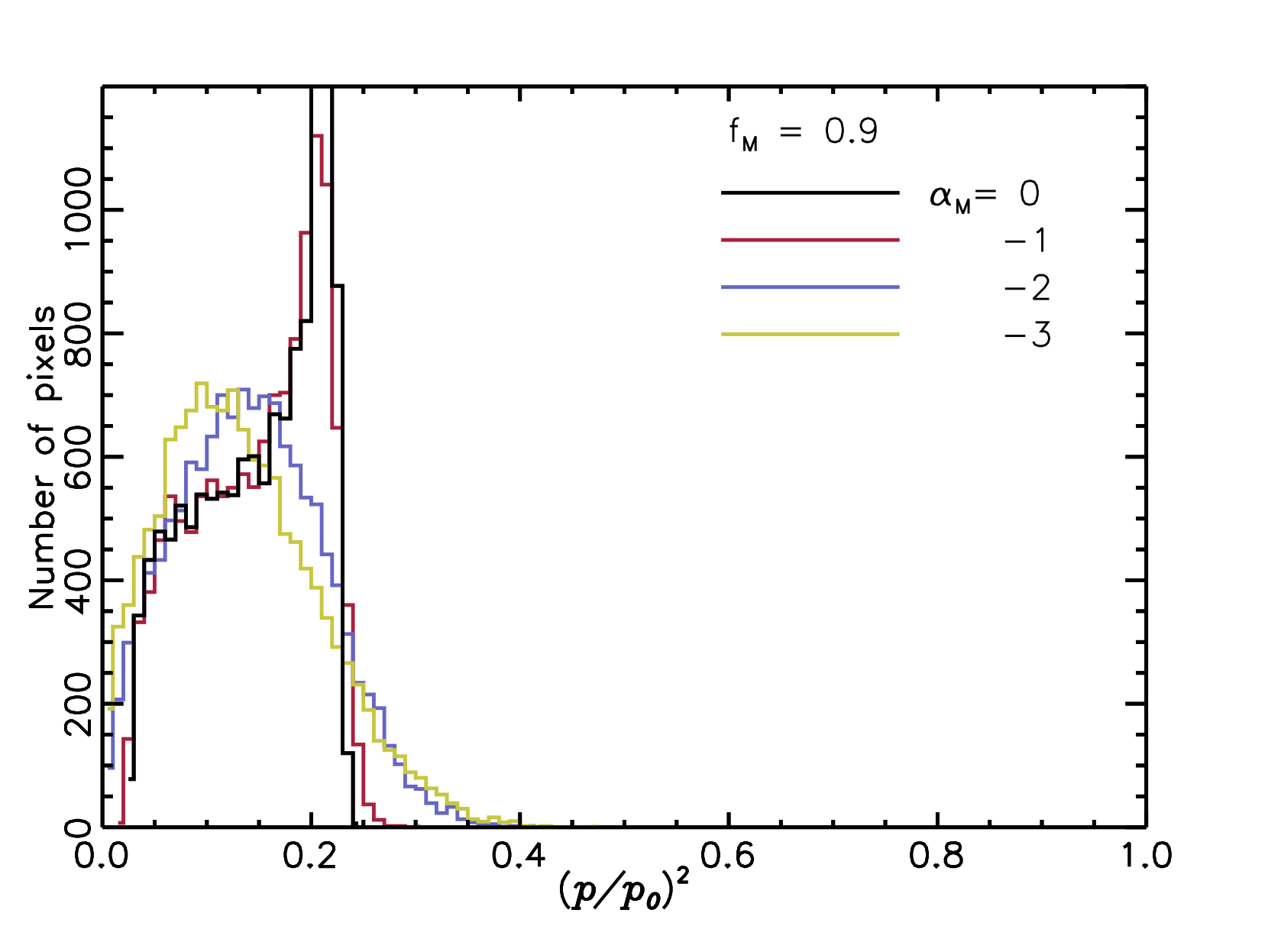}
\end{tabular}
\caption[]{Model histograms of $p^2$ normalized to unity with $p_0$, obtained
  for a continuously varying GMF orientation along the LOS, with
  $f_{\rm M}=0.9$ for several values of $\alpha$, between $0$ (black curve)
  and $-3$ (yellow curve). To facilitate the comparison of the histogram of
  $(p/p_0)^2$ with that in Fig.\ref{fig:Modelayers}, we have used the same bin
  width (0.01) to compute both histograms.}
\label{fig:slopes}
\end{figure}

In their modelling of dust polarization in molecular clouds, \citet{Myers91}
and \citet{planck2015-XXXV} introduced a correlation length that is
associated with the coupling scale through collisions between ions and
neutrals. For the \Planck\ data relating to the diffuse ISM, we propose a
different interpretation.
Following \citet{Eilek1989_2}, we derive the correlation length of the
turbulent component of the GMF ($l_{\rm c}$) from the 2-point
auto-correlation function, $C_{B}$, of each of the three components
of $B_{\rm t}$:
\begin{equation}\label{eq:eilek}
 \int C_{B}(s) \, \mathrm{d} s = l_{\rm c} \, \sigma_{B}^2,
\end{equation}
where $s$ is the lag of $C_{ B}$ along one given direction and
$\sigma_{B}$ is the dispersion of $B_i$. 
In this framework, the
number of correlation lengths along the LOS is $N_{\rm c}=L/l_{\rm c}$,
where $L$ is the effective extent of matter along the LOS. 
We compute $C_{B} $ from Gaussian realizations of $B_i$  for power-law spectra\footnote{To a good approximation, 
$\sigma_{B}^2 - C_{B}$ can be fitted with a power law of the lag $s$}, and, from there, $N_{\rm c}$   integrating
Eq.~(\ref{eq:eilek}) up to the lag where $C_{B}=0$.  
$N_{\rm c}$ depends on the spectral index $\alpha$ of the power spectrum of the components of $B_{\rm t}$. 
We find values of $N_{\rm c}$ of 16, 10, 6, and 5 for spectral indices of the power law spectrum $\alpha = -1.5$, $-2$, $-2.5$, and $-3$,
respectively. 

We can now compute the Stokes parameters for this continuous description of
$\vec{B}_{\rm t}$ and the mean orientation, $\vec{B}_0$ (determined in
Sect.~\ref{sec:ordfield}), through the integral equations described in
Appendix~\ref{app:approx}, for several values of $\alpha$ using a constant
source function as in step C. The Gaussian realizations and the integrals are
computed over 1024 of points along each LOS at $b < -60^\circ$.
In this approach, used earlier by \citet[e.g.,][]{MAMD2008,ODea12}, there is
no correlation of $\vec{B}_{\rm t}$ between
nearby pixels on the sky. Hence, we cannot produce realistic images but we
do sample the 1-point distribution of $p^2$. 

The histograms of $p^2$
(normalized to unity with $p_0$) are presented in Fig.~\ref{fig:slopes} for
several values of $\alpha$, with $f_{\rm M}=0.9$ and no data noise. We use
the same binning as in Fig.~\ref{fig:Modelayers} to allow
for a direct comparison between the two sets of histograms. 
The continuous description of $\vec{B}_{\rm t}$ matches the standard
deviation of $\psi_R$ measured in the \planck\ data for $\alpha\simeq-3$.
However, the corresponding histogram of $(p/p_0)^2$ in Fig.~\ref{fig:slopes}
is narrower than the one for $N=7$ in Fig.~\ref{fig:Modelayers}, which fits
the data better.  We conclude that the number of polarization layers may
be interpreted as the number of effective modes contributing 
to the variations of the orientation of $\vec{B}_{\rm t}$ along the LOS
within the WNM and WIM.  From this view point, the low value of $N$ derived from the data
fit reflects the steepness of the power spectrum of $\vec{B}_{\rm t}$;
however, this interpretation does not fully account for the data, because
it ignores the density structure of the diffuse ISM (i.e., the CNM).

\subsection{Future perspectives}\label{ssec:perspectives}

We now briefly outline a few future directions that could be taken to
extend our data analysis and modelling.

We have started to investigate the impact of the GMF structure on the statistics of the polarization parameters. In an
upcoming paper we will use the model presented in this work to reproduce
the dust polarization power spectra measured by \planck\
\citep{planck2014-XXX} and constrain the value of $\alpha_{\rm M}$,
the value of which is left open in this paper. 
Another future project will be to introduce the density structure and
its correlation with the orientation of the GMF within each polarization layer.
Such a study will enable us to assess the respective contributions of
the density and the GMF structure to the statistics of the dust
polarization data.

In the present work, we have aimed at providing a phenomenological method
to compute realizations of the dust polarization sky for 
component separation in measurements of the polarization
of the CMB. We want to stress the simplicity of our approach, which
allowed us to characterize the high latitude polarization sky with very few
parameters. This framework might be useful to predict the expected accuracy
of component-separation methods in future CMB experiments.
\citet{planck2015-XXXVIII} and \citet{Clark15} associated
the asymmetry between $EE$ and $BB$ power spectra of dust
polarization \citep[i.e., $C_\ell^{BB}\simeq 0.5 \, C_\ell^{EE}$,][]{planck2014-XXX}
with the correlation between the structure
of the GMF and the distribution of interstellar matter. 
Future models will need to take this correlation into account in order to
realistically assess the accuracy to which, for a given experiment,
dust and CMB polarization can be separated.


\section{Summary}\label{sec:summary} 

We have analysed the \planck\ maps
of the Stokes parameters at high Galactic latitudes over 
the sky area $b<-60^{\circ}$, which is well suited for describing the
Galactic magnetic field (GMF)
structure in the diffuse interstellar medium (ISM), and is 
directly relevant for cosmic microwave background (CMB) studies.
We characterized the structure of the Stokes parameter maps at 353\,GHz,
as well as the statistics of the polarization fraction $p$ and angle $\psi$. 
We presented simple geometrical models, which
relate the data to the structure of the GMF in the solar neighbourhood.
Combining models of the turbulent and ordered components of the GMF, we have
reproduced the patterns of the Stokes $Q$ and $U$ maps at large angular
scales, as well as the histograms of $p$ and $\psi$.
The main results of the paper are listed below.

\begin{itemize}

\item We find that the histogram of $p$ at high
  Galactic latitudes has a similar dispersion as that measured over the
  whole sky, although with a smaller depolarization, caused by line-of-sight
  (LOS) variations of the GMF orientation, on and near the Galactic plane. 

\item The Stokes $Q$ and $U$ maps show regular patterns at large
  scales, which we associate with the mean orientation of the GMF in the solar
  neighbourhood. We build a geometric model and find a mean
  orientation towards Galactic coordinates
  $(l_0,b_0)=(70^\circ,24^\circ)$, compatible with previous
  estimates. The fit also provides us with the average value of $p$ at
  $b \le -60^{\circ}$, which is  $(12 \pm 1)\,\%$. 

\item By means of a simple description of the turbulent component of
  the GMF (Gaussian and isotropic), we manage to 
  account for both the dispersion of $\psi$ and the
  histogram of $p$. The effect of depolarization caused by the GMF
  fluctuations along the LOS is introduced through an approximation where
  the integrals along the LOS are replaced by a discrete sum over only
  a few independent polarization layers.  This approach successfully
  reproduces the $p$ and $\psi$ distributions 
  using $N\simeq4$--9 layers. 
  
  \item
  The integration along the LOS  generates a mean depolarization factor that is about 0.5
  and thus leads to an estimate of $p_0$ about twice greater than the average value of $p$.
  The best-fit value of the effective polarization of dust, which
  combines the intrinsic polarization of dust grains and their degree
  of alignment with the GMF, is $(26 \pm 3)\,\%$.

\item Our description of the turbulent component of the GMF corresponds to a
  rough equality between the turbulent and mean strengths of the GMF. The
  same conclusion was reached from modelling the dispersion of polarization
  angles measured for CNM filamentary structures by \citet{planck2014-XXXII}. 
  We extend this to the diffuse ISM observed in the high latitude sky, which
  comprises of both WNM and CNM gas.
 
\end{itemize}

The present study represents the first step towards the characterization of the
magnetized properties of the diffuse ISM by means of the \planck\ data. 
We argue that both the density structure and the effective correlation
length of the GMF contribute to account for the large dispersion of $p$
observed in the data. This can be further investigated using MHD numerical
simulations.  The next step in our modelling of dust polarization at high
Galactic latitudes will be to fit the $E$ and $B$ power spectra. 
This will constrain the spectral index
of the GMF power spectrum, providing information on the
turbulent energy cascade in the diffuse ISM. It is also a required step
before using our model to compute simulated maps for assessing
component-separation methods in CMB polarization projects.

\begin{acknowledgements}
The Planck Collaboration acknowledges the support of: ESA; CNES, and
CNRS/INSU-IN2P3-INP (France); ASI, CNR, and INAF (Italy); NASA and DoE
(USA); STFC and UKSA (UK); CSIC, MINECO, JA, and RES (Spain); Tekes, AoF,
and CSC (Finland); DLR and MPG (Germany); CSA (Canada); DTU Space
(Denmark); SER/SSO (Switzerland); RCN (Norway); SFI (Ireland);
FCT/MCTES (Portugal); ERC and PRACE (EU). A description of the Planck
Collaboration and a list of its members, indicating which technical
or scientific activities they have been involved in, can be found at
{\tt http://www.cosmos.esa.int/web/planck/planck{\-}collaboration}.
The research leading to these results has received funding from the European
Research Council under the European Union's Seventh Framework Programme
(FP7/2007-2013) / ERC grant agreement No.~267934.
\end{acknowledgements}

\bibliographystyle{aa}
\def\eprinttmppp@#1arXiv:@{#1}
\providecommand{\arxivlink[1]}{\href{http://arxiv.org/abs/#1}{arXiv:#1}}
\def\eprinttmp@#1arXiv:#2 [#3]#4@{\ifthenelse{\equal{#3}{x}}{\ifthenelse{
\equal{#1}{}}{\arxivlink{\eprinttmppp@#2@}}{\arxivlink{#1}}}{\arxivlink{#2}
  [#3]}}
\providecommand{\eprintlink}[1]{\eprinttmp@#1arXiv: [x]@}
\providecommand{\eprint}[1]{\eprintlink{#1}}
\providecommand{\adsurl}[1]{\href{#1}{ADS}}

\appendix

\section{Approximations for dust polarization}\label{app:approx}

In this Appendix, we detail the approximations made to model the Stokes
parameters for linear polarization from dust emission. For
the sake of clarity, we recall the integral equations of the Stokes
parameters $I$, $Q$, and $U$ from \citet{planck2014-XX}:
\begin{subequations}
\begin{align}
\label{eq:simulated_I}
\StokesI &=\int S_\nu\,e^{-\tau_\nu}\left[1-p_{0}
 \left(\cos^2\polangsky-\frac{2}{3}\right)\right]\mathrm{d}\tau_\nu;\\
\label{eq:simulated_Q}
\StokesQ&=\int p_{0}\,S_\nu\,e^{-\tau_\nu}\cos\left(2\psi\right)
 \cos^2\polangsky\,\mathrm{d}\tau_\nu;\\
\label{eq:simulated_U}
\StokesU&=\int p_{0}\,S_\nu\,e^{-\tau_\nu}\sin\left(2\psi\right)
 \cos^2\polangsky\,\mathrm{d}\tau_\nu.
\end{align}
\end{subequations} 
Here $\tau_\nu$ is the optical depth and $S_{\nu}$ is the source
function of dust emission, while $p_{0}$ and the angles ($\psi,\gamma$) are
the same as in Sect.~\ref{ssec:Stokespara}. 

We make two additional points:
(1) in order to relate $p$ as shown in Eq.~(\ref{eq:p_obs}) to the mean
orientation of the GMF with respect to the POS (the angle $\gamma$), we need
to assume that all parameters in Eqs.~(\ref{eq:simulated_I}),
(\ref{eq:simulated_Q}), and (\ref{eq:simulated_U}) are roughly uniform along
the LOS; and 
(2) the total intensity in Eq.~(\ref{eq:simulated_I}) also depends
on the GMF orientation through the angle $\gamma$. However, throughout our
modelling procedure, we neglect this dependence.

In general the corrections to Stokes $I$ caused by the GMF geometry are
small, ranging roughly
range between $-7\,\%$ and $+13\,\%$ for $p_0\simeq20\,\%$
\citep{planck2014-XIX}. In our study, we focus on a region of the sky
where the depolarization produced by $\cos^2\gamma$ is small ($\cos^2\gamma$
is mostly close to unity over the southern Galactic cap).
Hence, in our study, the correction to Eq.~(\ref{eq:simulated_I})
would always be negative and less than $10\,\%$. Thus, in
Sect.~\ref{ssec:levels} we might estimate a value of
$p_0$ slightly greater than the true value that we would have obtained
by modelling the GMF correction for Stokes $I$. In practice
Eq.~(\ref{eq:stokesAfit}) in Sect.~\ref{ssec:modelA2} would change as follows:
\begin{align}
\label{eq:stokesAfitcor}
& Q_{353}= \frac{p_0q_{\rm A}}{1-p_0(\cos^2\gamma
 - \frac{2}{3})}D_{353};\nonumber \\
& U_{353}= \frac{p_0u_{\rm A}}{1-p_0(\cos^2\gamma
 - \frac{2}{3})}D_{353}.
\end{align}
The fits of steps A, B, and C would then not be linear in $p_0$
anymore, substantially complicating the fit.  We argue that, considering
the overall approximations (analytical and astrophysical) of our
models, the GMF geometry in Stokes $I$ is a minor issue. 

\raggedright

\end{document}